# COMPUTER MODEL VALIDATION WITH FUNCTIONAL OUTPUT[1]

By M. J. Bayarri, J. O. Berger, J. Cafeo, G. Garcia-Donato,
F. Liu, J. Palomo, R. J. Parthasarathy, R. Paulo,
J. Sacks and D. Walsh

*Universitat de Valencia, Duke University, General Motors, Universidad de
Castilla-La Mancha, Duke University, Universidad Rey Juan Carlos,
General Motors, ISEG Technical University of Lisbon, National Institute
of Statistical Sciences and Massey University*

A key question in evaluation of computer models is *Does the
computer model adequately represent reality?* A six-step process for
computer model validation is set out in Bayarri et al. [*Technometrics*
**49** (2007) 138–154] (and briefly summarized below), based on com-
parison of computer model runs with field data of the process being
modeled. The methodology is particularly suited to treating the ma-
jor issues associated with the validation process: quantifying multiple
sources of error and uncertainty in computer models; combining mul-
tiple sources of information; and being able to adapt to different, but
related scenarios.

Two complications that frequently arise in practice are the need to
deal with highly irregular functional data and the need to acknowl-
edge and incorporate uncertainty in the inputs. We develop method-
ology to deal with both complications. A key part of the approach
utilizes a wavelet representation of the functional data, applies a hier-
archical version of the scalar validation methodology to the wavelet
coefficients, and transforms back, to ultimately compare computer
model output with field output. The generality of the methodology
is only limited by the capability of a combination of computational
tools and the appropriateness of decompositions of the sort (wavelets)
employed here.

Received February 2006; revised December 2006.
[1]Supported in part by NSF Grants DMS-00-73952 to the National Institute of Statis-
tical Sciences and DMS-01-03265 and Spanish Ministry of Education Grant MTM2004-
03290.
*AMS 2000 subject classification.* 62Q05.
*Key words and phrases.* Computer models, validation, functional data, bias, Bayesian
analysis, uncertain inputs.

The methods and analyses we present are illustrated with a test bed dynamic stress analysis for a particular engineering system.

## 1. Introduction.

1.1. *Validation framework.* Bayarri et al. [4] described a general framework for validation of complex computer models and applied the framework to two examples involving scalar data. The framework consists of a six-step procedure to treat the combined calibration/validation process, and to assess the possible systematic differences between model outcomes and test outcomes (so-termed biases), by estimating these biases along with uncertainty bounds for these estimates. The six steps are (1) defining the problem (inputs, outputs, initial uncertainties); (2) establishing evaluation criteria; (3) designing experiments; (4) approximating computer model output; (5) analyzing the combination of field and computer run data; (6) feeding back to revise the model, perform additional experiments, and so on. Bayarri et al. [3] generalized this work to the situation of smooth functional data arising within a hierarchical structure.

This paper is an extension of Bayarri et al. [4], motivated by methodological needs in analyzing a computer model for analyzing stress on an engineering system, a vehicle suspension system, subject to forces over time (the test bed problem is described in the following section). This involves the following important—and technically challenging—problems that greatly widen the applicability of the six-step strategy.

PROBLEM 1 (*Irregular functional output*). In Bayarri et al. [4], examples involved real-valued model outputs; in Bayarri et al. [3], outputs were smooth functions that could be handled by the simple device of including the function argument (time) as an input of the system. In many engineering scenarios with functional output, functions are not smooth, and adding the function variable to the list of inputs can result in a computationally intractable problem. This is so in the test bed problem, for instance, with typical irregular functional data indicated in Figure 1.

PROBLEM 2 (*Uncertainty in inputs*). A second ubiquitous problem in engineering scenarios is that (unmeasured) manufacturing variations are present in tested components; incorporating this uncertainty into the analysis can be crucial.

PROBLEM 3 (*Prediction in altered or new settings*). The point of computer modeling in engineering contexts is typically to allow use of the computer model to predict outcomes in altered or new settings for which no field data are available. We consider several approaches to this problem.

1.2. *Background and motivation.* General discussions of the entire validation and verification process can be found in Roache [22], Oberkampf and Trucano [20], Cafeo and Cavendish [5], Easterling [9], Pilch et al. [21], Trucano, Pilch and Oberkampf [26] and Santner, Williams and Notz [24]. We focus here on the last stage of the process: assessing the accuracy of the computer model in predicting reality, and using both the computer model and field data to make predictions, especially in new situations.

Because a computer model can virtually never be said to be a completely accurate representation of the real process being modeled, the relevant question is "Does the model provide predictions that are accurate enough for the intended use of the model?" Thus predictions need to come with what were called *tolerance bounds* in Bayarri et al. [4], indicating the magnitude of the possible error in prediction. This focus on giving tolerance bounds, rather than stating a yes/no answer as to model validity, is important for several reasons: (i) Models rarely give highly accurate predictions over the entire range of inputs of possible interest, and it is often difficult to characterize regions of accuracy and inaccuracy; (ii) The degree of accuracy that is needed can vary from one application of the computer model to another; (iii) Tolerance bounds account for *model bias*, the principal symptom of model inadequacy–accuracy of the model cannot simply be represented by a variance or standard error.

The key components of the approach outlined here are the use of Gaussian process response-surface approximations to a computer model, following on work in Sacks et al. [23], Currin et al. [8], Welch et al. [28] and Morris, Mitchell and Ylvisaker [16], and introduction of Bayesian representations of model bias and uncertainty, following on work in Kennedy and O'Hagan [14] and Kennedy, O'Hagan and Higgins [15]. A related approach to Bayesian analysis of computer models is that of Craig et al. [7], Craig et al. [6] and Goldstein and Rougier [11, 12], which focus on utilization of linear Bayes methodology.

1.3. *Overview.* Section 2 describes the test bed example and associated data. Section 3 has the formulation of the statistical problem and assumptions that we make for the analysis. In Section 4 we set down the methods of analysis with the results in Section 5. Most of the details of computations are relegated to the Appendices.

**2. The test bed.** The test bed case study is about predicting loads resulting from stressful events on a vehicle suspension system over time, for example, hitting a pothole. In the initial part of the study there are seven unmeasured parameters of the system with specified nominal (mean) values $\mathbf{x}_{\text{nom}}$ (referred to later on as Condition A) and unknown manufacturing

TABLE 1

*I/U Map. Uncertainty ranges for the two calibration parameters and the seven input parameters that are subject to manufacturing variation*

| Parameter | Type | Uncertainty range |
|-----------|------|-------------------|
| $Damping_1$ | Calibration | [0.125, 0.875] |
| $Damping_2$ | Calibration | [0.125, 0.875] |
| $x_1$ | Nominal + Variation | [0.1667, 0.8333] |
| $x_2$ | Nominal + Variation | [0.1667, 0.8333] |
| $x_3$ | Nominal + Variation | [0.2083, 0.7917] |
| $x_4$ | Nominal + Variation | [0.1923, 0.8077] |
| $x_5$ | Nominal + Variation | [0.3529, 0.6471] |
| $x_6$ | Nominal + Variation | [0.1471, 0.8529] |
| $x_7$ | Nominal + Variation | [0.1923, 0.8077] |

variations $\boldsymbol{\delta}^*$. There are other relevant parameters that are known and fixed and hence not part of the experiments.

Field data are obtained by driving a vehicle over a proving ground course and recording the time history of load at sites on the suspension system. The curves must be registered (Appendix A) to assure that peaks and valleys occur at the same place.

In addition, there is a computer model aimed at producing the same response. The computer model is a so-termed ADAMS model, a commercially available, widely used finite-element based code that analyzes complex dynamic behavior (e.g., vibration, stress) of mechanical assemblies. The computer model has within it two calibration parameters $\mathbf{u}^* = (u_1^*, u_2^*)$ quantifying two different types of damping (unknown levels of energy dissipation) that need to be estimated (or tuned) to produce a matching response.

For proprietary reasons the specific parameters are not fully described—they include characteristics of tires, bushings and bumpers as well as vehicle mass. In addition, the values assigned to these parameters are coded on a [0, 1] scale and the output responses are also coded. In the coded scale the fixed values of $\mathbf{x}_{\text{nom}}$ are all 0.50. The uncertainty ranges for the nine parameters were elicited through extensive discussion with engineers and modelers; they are given in Table 1, the so-termed Input/Uncertainty (I/U) map (Bayarri et al. [4]). Along with the ranges, prior distributions were elicited for $(\mathbf{u}^*, \boldsymbol{\delta}^*)$ in Section 3.4.

*Field data.* In the initial study with Condition A inputs, a single vehicle was driven over a proving ground course seven times. The recorded field data consist of the time history of load at two sites on the suspension system. Plots of the output for Site 1 can be seen in Figure 1 for two of the time periods of particular interest. Thus there are seven replicates and a single $\mathbf{x}_{\text{nom}}$ in the field data.

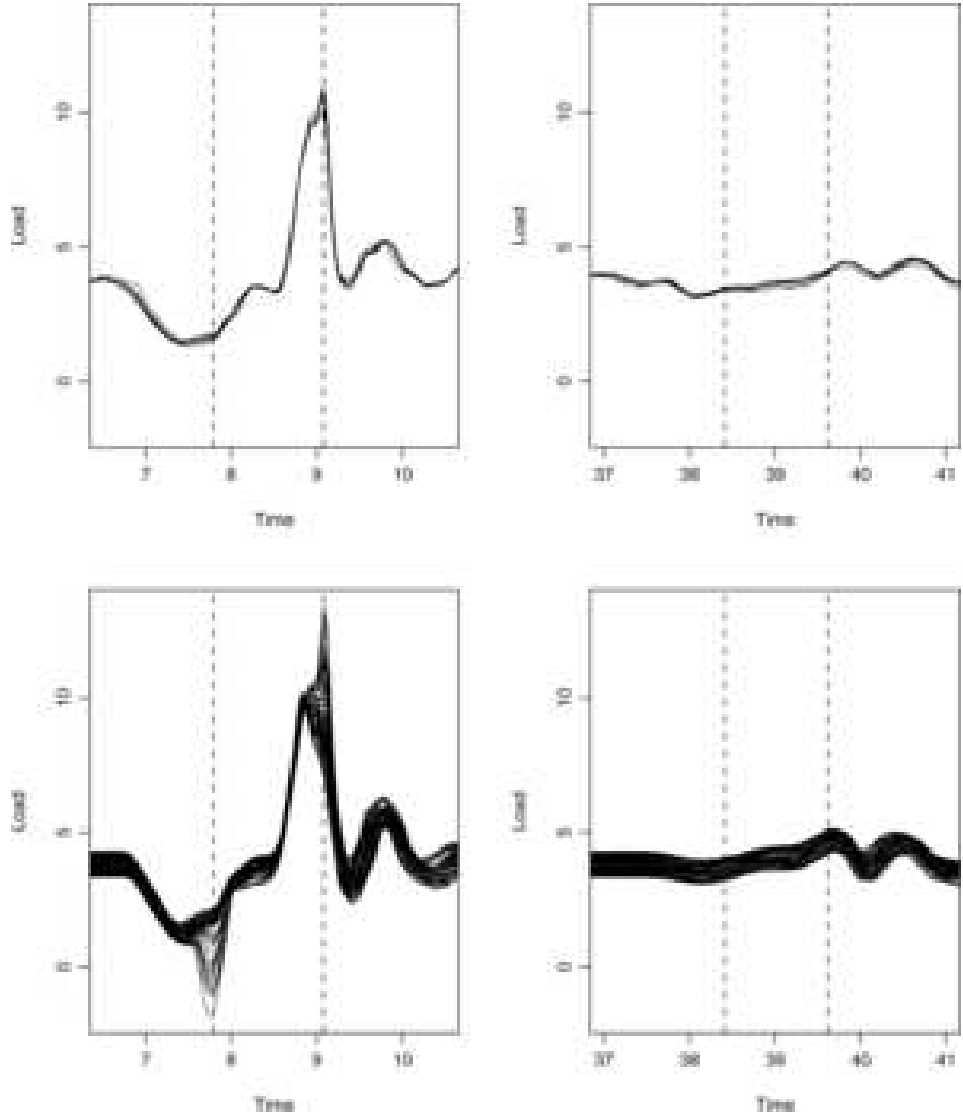

FIG. 1. *Model output (bottom) and registered field output (top) for Site 1 at Region 1 (left) and Region 2 (right). Vertical lines indicate the reference peak locations used in the curve registration discussed in Appendix* A.

*Computer model runs.* A typical model run for the test bed example takes one hour, limiting the number of runs that can feasibly be made. To select which runs to make we adopted the design strategy used in Bayarri et al. [4]: the 9-dimensional rectangle defined by the ranges of the parameters in Table 1 is first transformed into the 9-dimensional unit cube and a 65-point Latin hypercube design (LHD) is then selected using code by W. Welch that finds an approximately maximin LHD. The actual design points, as well as a 2-dimensional projection of the design, can be found in Bayarri et al. [2]. In addition, we added a point at the center $(0.5, \ldots, 0.5)$, the nominal values. One run failed to converge and was deleted from the experiment leaving a total of 65 design points.

## 3. Formulation, statistical model and assumptions.

3.1. *Formulation.* Given a vector of inputs, $\mathbf{x} = (x_1, \ldots, x_d)$, to a time-dependent system, denote the "real" response over time $t$ as $y^R(\mathbf{x}; t)$. Field measurement of the real response has error and we write the $r$th replicate field measurement as

$$y_r^F(\mathbf{x}; t) = y^R(\mathbf{x}; t) + \varepsilon_r(t),\tag{1}$$

where the $\varepsilon_r(\cdot)$'s are independent mean zero Gaussian processes (described below). Some inputs may also have error; we take that into account below.

In addition, there is a computer model aimed at producing the same response. The computer model may have within it calibration/tuning parameters $\mathbf{u} = (u_1, \ldots, u_m)$ that need to be estimated (or tuned) to produce a matching response. The model output is then of the form $y^M(\mathbf{x}, \mathbf{u}; t)$; it is affected by $\mathbf{u}$ but the real response, $y^R$, is not. The connection between model output and reality is then expressed in

$$y^R(\mathbf{x}; t) = y^M(\mathbf{x}, \mathbf{u}^*; t) + b(\mathbf{x}; t),\tag{2}$$

where $\mathbf{u}^*$ is the true value of the (vector) calibration parameter, $y^M(\mathbf{x}, \mathbf{u}^*; t)$ is the model response at time $t$ and the true value of $\mathbf{u}$, and $b(\mathbf{x}; t)$, defined by subtraction, is the associated bias. In situations where $\mathbf{u}$ is a tuning parameter, there is no "true value" so $\mathbf{u}^*$ should be thought of as some type of fitted value of $\mathbf{u}$, with the bias defined relative to it.

Data from the field and from the computer model runs provide information for estimating the unknowns in (1) and (2). The Bayesian analysis we employ takes note of the fact, as in Bayarri et al. [4], that the unknowns $\mathbf{u}^*$ and the bias are not statistically identifiable and, consequently, specification of their prior distributions is of particular importance for the analysis.

*Inputs.* Some inputs may be specified or physically measured with essentially perfect accuracy. Those that remain fixed for both field data and model runs play no further role and are not part of $\mathbf{x}$. Other (unmeasured) inputs will have specified nominal values (generally, they will vary in the experiments) that are subject to manufacturing variation with specified distributions. We write these as

$$\mathbf{x} = \mathbf{x}_{\mathrm{nom}} + \boldsymbol{\delta},\tag{3}$$

where $\mathbf{x}_{\mathrm{nom}}$ is the known nominal value and the distribution of the manufacturing variation $\boldsymbol{\delta}$ can be specified. In effect this transforms (2) into

$$y^R(\mathbf{x}_{\mathrm{nom}} + \boldsymbol{\delta}^*; t) = y^M(\mathbf{x}_{\mathrm{nom}} + \boldsymbol{\delta}^*, \mathbf{u}^*; t) + b(\mathbf{x}_{\mathrm{nom}} + \boldsymbol{\delta}^*; t),$$

where $\boldsymbol{\delta}^*$ is the actual (unknown) value of $\boldsymbol{\delta}$. The parameters $\boldsymbol{\delta}$ are like calibration parameters in that they are unknown but physically real.

*The unknowns.* Prior to making computer runs or collecting field data, the unknowns in (1) and (2) are $(y^M, \mathbf{u}^*, \boldsymbol{\delta}^*, b, V_\epsilon)$, where $V_\epsilon$ is the covariance function of $\epsilon$. A full Bayesian analysis would contemplate placing priors on these unknowns and, given field data and model runs, produce posterior distributions. But the complexity (e.g., of irregular functional output) and high dimensionality militate against such a strategy unless simplifications can be made. One such is the use of a basis representation of the functional data. In particular, to handle the irregular functions, we will consider wavelet decompositions. Other settings may allow different representations such as Fourier series or principal components (Higdon et al. [13]).

3.2. *Wavelet decomposition.* The nature of the functions in Figure 1, for example, suggests that wavelet decomposition would be a suitable basis representation (see Vidakovic [27]; Müller and Vidakovic [18] and Morris et al. [17] are among other references with applications related to ours).

The wavelet decomposition (more details are in Appendix A) we use for $y^M$ is of the form

$$(4) \qquad y^M(\mathbf{x}, \mathbf{u}; t) = \sum_i w_i^M(\mathbf{x}, \mathbf{u}) \Psi_i(t),$$

where the wavelet basis functions $\Psi_i(t)$ are default choices in *R wavethresh* (Daubechies wavelets of index 2; for simplicity of notation, we include the scaling function as one of the basis elements). Similarly, the field curves ($r$th replicate) are represented as

$$(5) \qquad y_r^F(\mathbf{x}; t) = \sum_i w_{ir}^F(\mathbf{x}) \Psi_i(t).$$

A thresholding procedure, used to produce a manageable number of coefficients while maintaining adequate accuracy, leads to the approximations

$$\begin{aligned} (6) \qquad y^M(\mathbf{x}, \mathbf{u}; t) &= \sum_{i \in I} w_i^M(\mathbf{x}, \mathbf{u}) \Psi_i(t), \\ y_r^F(\mathbf{x}; t) &= \sum_{i \in I} w_{ir}^F(\mathbf{x}) \Psi_i(t). \end{aligned}$$

The accuracy of the approximations using the reduced set of elements for the test bed problem is indicated in Figure 2. Since similarly accurate reconstruction resulted for all field and model run curves, we also assume that

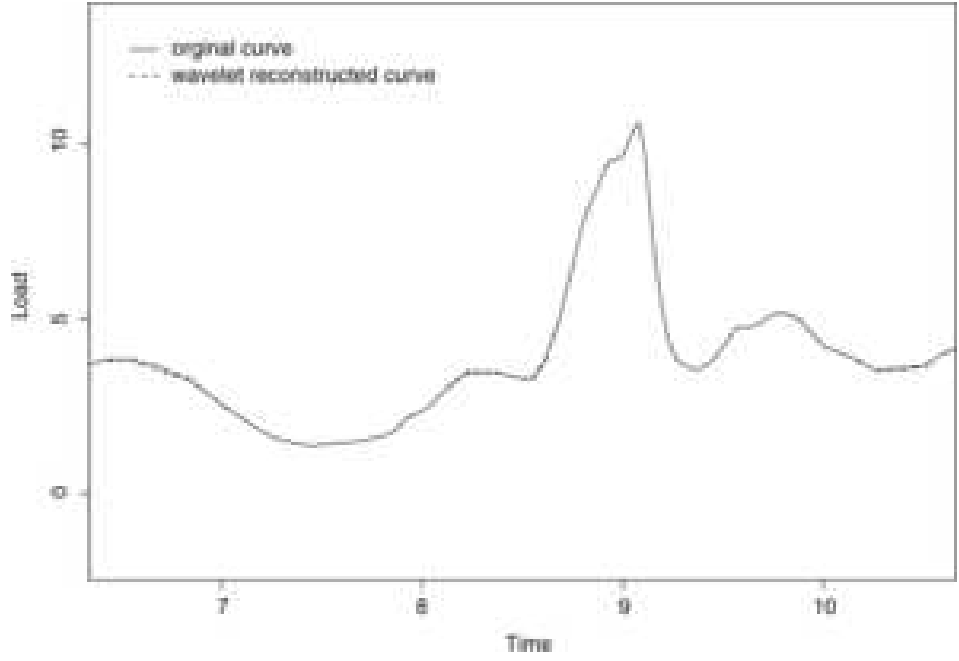

Fig. 2.   *The original curve and wavelet reconstructed curve for the first field-run at Site 1.*

reality and the bias function can be accurately represented by the same basis elements, and write

$$y^R(\mathbf{x}; t) = \sum_{i \in I} w_i^R(\mathbf{x}) \Psi_i(t),$$

(7)
$$b(\mathbf{x}; t) = \sum_{i \in I} w_i^b(\mathbf{x}) \Psi_i(t),$$

$$\varepsilon_r(t) = \sum_{i \in I} \varepsilon_{ir} \Psi_i(t).$$

Matching coefficients and using (1) and (2), we get

(8)   $$w_i^R(\mathbf{x}) = w_i^M(\mathbf{x}, \mathbf{u}^*) + w_i^b(\mathbf{x}) \qquad \forall i \in I,$$

(9)   $$w_{ir}^F(\mathbf{x}) = w_i^R(\mathbf{x}) + \varepsilon_{ir} \qquad\qquad \forall i \in I.$$

That the measurement errors in the wavelet domain, $\varepsilon_{ir}$, are normally distributed with mean zero and are independent across replications $r$ follows from the Gaussian process assumption on the process measurement error. We will, in addition, assume that these coefficients are independent across $i$, with possibly differing variances $\sigma_i^2$. This independence assumption is imposed so that later computations are feasible. That might seem unrealistic, given that the seven residual functions $(y_r^F(\mathbf{x}; t) - \bar{y}_r^F(\mathbf{x}; t))$—shown in the top of Figure 3—can be seen to be correlated in time $t$, suggesting that the $\varepsilon_{ir}$ should be correlated in $i$. But, as long as the $\sigma_i^2$ differ, independent normal $\varepsilon_{ir}$ lead to a Gaussian process with mean zero and covariance function $\sum_{i \in I} \sigma_i^2 \Psi_i(t) \Psi_i(t')$. The bottom of Figure 3 gives seven realizations of this process, with the $\sigma_i^2$ being estimated by the usual unbiased estimates, based on the replicates. The correlation patterns of the two processes seem quite similar.

Our approach is to analyze each of the retained wavelet coefficients in (8) and (9), and recombine them to obtain estimates and uncertainties for the "original" functions in (1) and (2).

3.3. *GASP approximation.* For $y^M$, the wavelet coefficients are functions of $(\mathbf{x}, \mathbf{u})$. Because we cannot freely run the computer model for every $(\mathbf{x}, \mathbf{u})$, we approximate each of the retained coefficients using data from computer runs. Formally, we start with a Gaussian process prior distribution on a coefficient $w_i^M(\mathbf{x}, \mathbf{u})$. Given computer model runs, $y^M(\mathbf{x}_k, \mathbf{u}_k)$, where $\{(\mathbf{x}_k, \mathbf{u}_k), k = 1, \ldots, K\}$ are the design points in a computer experiment, we extract the data $\{w_i^M(\mathbf{x}_k, \mathbf{u}_k)\}$ and approximate $w_i^M(\mathbf{x}, \mathbf{u})$ as the Bayes predictor, the posterior mean, of $w_i^M(\mathbf{x}, \mathbf{u})$ given the data.

The Gaussian process priors we use are as in the GASP methodology described in Bayarri et al. [4]. Let $\mathbf{z} = (\mathbf{x}, \mathbf{u})$. For each $i \in I$ (the set of retained wavelet coefficients), the GASP assumption is that $w_i^M(\mathbf{z})$ is a Gaussian process with mean $\mu_i^M$, constant variance $1/\lambda_i^M$ and correlation function

$$c_i^M(\mathbf{z}, \mathbf{z}') = \exp\left(-\sum_{p=1}^{n_M} \beta_{ip}^M |z_p - z_p'|^{2-\alpha_{ip}^M}\right),$$

where $n_M$ is the number of coordinates in $\mathbf{z}$, the $\beta$'s are nonnegative parameters and the $\alpha$'s are between 0 and 1.

Let $\boldsymbol{\theta}_i^M = \{\mu_i^M, \lambda_i^M, \alpha_{ip}^M, \beta_{ip}^M; p = 1, \ldots, n_M\}$ be the collection of the (hyper) parameters determining the Gaussian prior distribution of $w_i^M$. To produce the Bayes predictor we have to deal with the $\boldsymbol{\theta}_i^M$'s; we do so in Section 4.

3.4. *Other prior specifications.* Priors for $\mathbf{u}^*, \boldsymbol{\delta}^*$ are context specific. Engineering advice led to adopting uniform priors for $\mathbf{u}^*$ on their ranges in Table 1. For the manufacturing variations of the unmeasured parameters in Table 1, the advice led to normal priors with standard deviations equal to 1/6 of the ranges of the uncertainty intervals in Table 1. Specifically,

$$\pi(u_1) = \pi(u_2) = \text{ Uniform on } [0.125, 0.875],$$

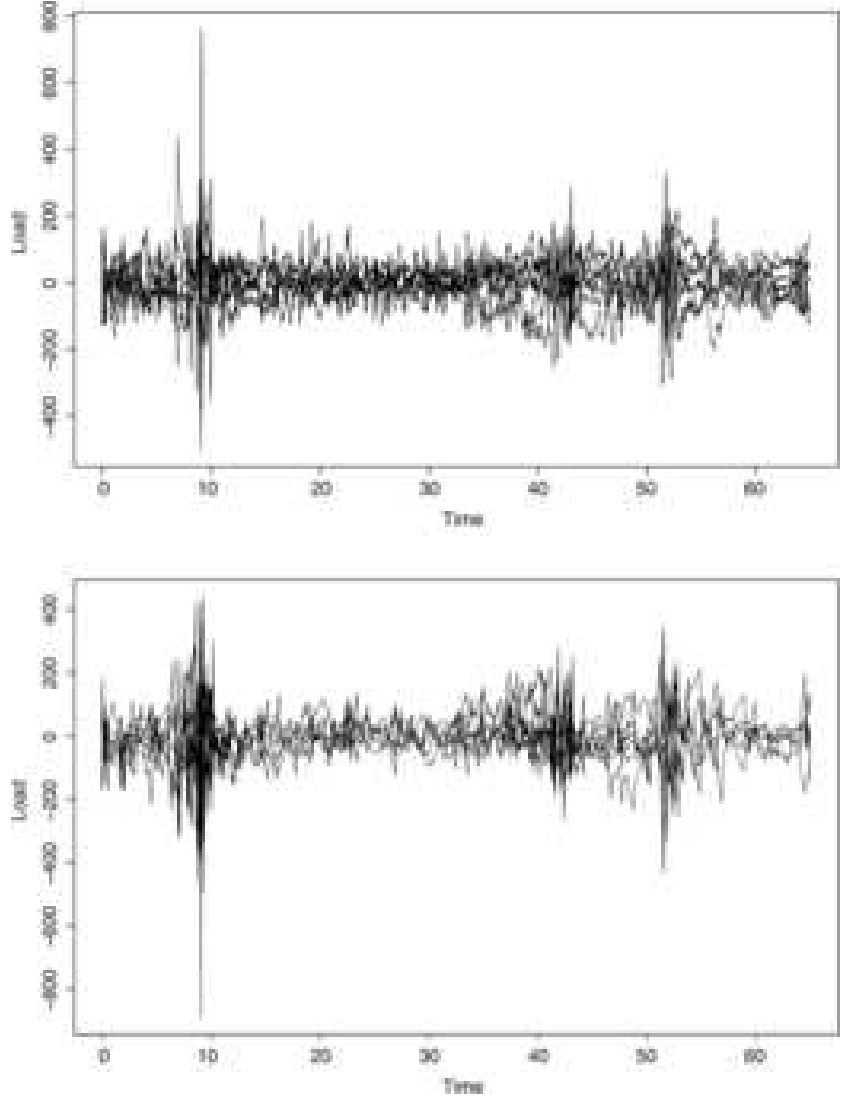

FIG. 3. *The seven residual field processes (top) and seven simulated error processes (bottom).*

$$\pi(\delta_1) = \pi(\delta_2) \sim N(0, 0.1111^2) \text{ truncated to } [-0.3333, 0.3333],$$

$$\pi(\delta_3) \sim N(0, 0.09723^2) \text{ truncated to } [-0.2917, 0.2917],$$

$$\pi(\delta_4) = \pi(\delta_7) \sim N(0, 0.1026^2) \text{ truncated to } [-0.3077, 0.3077],$$

$$\pi(\delta_5) \sim N(0, 0.04903^2) \text{ truncated to } [-0.1471, 0.1471],$$

$$\pi(\delta_6) \sim N(0, 0.1176^2) \text{ truncated to } [-0.3529, 0.3529].$$

The $\sigma_i^2$ are given the usual noninformative priors $\pi(\sigma_i^2) \propto 1/\sigma_i^2$. Priors for the bias wavelet coefficients, the $(w_i^b)$'s, will be Gaussian but restricted to depend only on those coordinates of $\mathbf{x}_{\text{nom}}$ that vary in the experiments. Because there is only one $\mathbf{x}_{\text{nom}}$ in the test bed experiment, we only consider the case where the $(w_i^b)$'s are constants, though a similar approach can be taken for more general settings.

In the wavelet decomposition, each $i \in I$ belongs to some resolution level, $j(i)$ (in the test bed—see Appendix A—the levels go from 0 to 12). It is natural and common to model wavelet parameters hierarchically, according to their resolution level. The priors for the biases $w_i^b$ at resolution level $j$ are

$$\pi(w_i^b \mid \tau_j^2) \sim N(0, \tau_j^2), \qquad j = 0, \ldots, 12. \tag{10}$$

This is a strong shrinkage prior, shrinking the biases to zero. One might be concerned with such strong shrinkage to zero, but the computer modeling world is one in which biases are typically assumed to be zero, so that utilizing a strong shrinkage prior has the appeal that any detected bias is more likely to be believed to be real in the community than would bias detected with a weaker assumption. (Of course, there are also statistical arguments for using such shrinkage priors.)

The hypervariances $\tau_j^2$ are assigned a variant of a typical objective prior for hypervariances,

$$\pi(\tau_j^2 \mid \{\sigma_i^2\}) \propto \frac{1}{\tau_j^2 + 1/7\bar{\sigma}_j^2},$$

where $\bar{\sigma}_j^2 =$ average of $\sigma_i^2$ for $i$ at level $j$. The $\bar{\sigma}_j^2$ provide a "scale" for the $\tau_j^2$ and are necessary—or at least some constants in the denominators are necessary—to yield a proper posterior.

**4. Estimation and analysis.** We restrict attention for the most part to the context of the test bed. It will be clear that much of what is done can be generalized.

*Approximating wavelet coefficients.* In Appendix A we find that, for the test bed, there are 289 wavelet coefficients $w_i^M$ to be treated. The Gaussian prior for each has 20 hyperparameters (coordinates of $\boldsymbol{\theta}_i^M$). A full Bayesian treatment would then require treatment of 5780 parameters, an infeasible task. Instead, we treat each coefficient separately and estimate each $\boldsymbol{\theta}_i^M$ by maximum likelihood based on the model-run data $\mathbf{w}_i^M = \{w_i^M(\mathbf{z}_k)\}$, using code developed by W. Welch. Recall that $\mathbf{z} = (\mathbf{x}, \mathbf{u})$ and denote the $k$th design point in the computer experiment by $\mathbf{z}_k$. For the test bed there are 65 $\mathbf{z}_k$'s.

Letting $\widehat{\boldsymbol{\theta}}_i^M = \{\widehat{\mu}_i^M, \widehat{\lambda}_i^M, \widehat{\alpha}_{ip}^M, \widehat{\beta}_{ip}^M; p = 1, \ldots, n_M\}$ be the maximum likelihood estimates of the $\boldsymbol{\theta}$'s, it follows that the GASP predictive distribution of $w_i^M(\mathbf{z})$ at a new $\mathbf{z}$ is

(11) $$w_i^M(\mathbf{z}) \mid \mathbf{w}_i^M, \widehat{\boldsymbol{\theta}}_i^M \sim N(\widehat{m}_i^M(\mathbf{z}), \widehat{V}_i^M(\mathbf{z})),$$

where

$$\widehat{m}_i^M(\mathbf{z}) = \widehat{\mu}_i^M + \widehat{\gamma}_i^M(\mathbf{z})'(\widehat{\boldsymbol{\Gamma}}_i^M)^{-1}(\mathbf{w}_i^M - \widehat{\mu}_i^M \mathbf{1}),$$

$$\widehat{V}_i^M(\mathbf{z}) = \frac{1}{\widehat{\lambda}_i^M} - \widehat{\gamma}_i^M(\mathbf{z})'(\widehat{\boldsymbol{\Gamma}}_i^M)^{-1}\widehat{\gamma}_i^M(\mathbf{z}),$$

with $\mathbf{1}$ the vector of ones, $\widehat{\boldsymbol{\Gamma}}_i^M$ ($65 \times 65$ in the test bed example) the covariance matrix for the model-run data $\mathbf{w}_i^M$ estimated by plugging in $\widehat{\boldsymbol{\theta}}_i^M$, and

$$\widehat{\gamma}_i^M(\mathbf{z}) = (1/\widehat{\lambda}_i^M)(\widehat{c}_i^M(\mathbf{z}_1, \mathbf{z}), \ldots, \widehat{c}_i^M(\mathbf{z}_k, \mathbf{z}))',$$

where $\widehat{c}$ is the estimated correlation function. For the rest of the paper we will use (11) as the definition of the GASP predictive distribution.

A plot of the $(\widehat{\alpha}, \widehat{\beta})$ pairs can be found in Bayarri et al. [2]. A $\beta$ near zero corresponds to a correlation near one and a function that is quite flat in that variable; an $\alpha$ near zero corresponds to a power of two in the exponent of the correlation, suggesting smooth functional dependence on that variable. For most of the pairs $(\widehat{\alpha}, \widehat{\beta})$, one or the other is near zero.

Full justification of the use of the plug-in maximum likelihood estimates for the parameters $\boldsymbol{\theta}_i^M$ is an open theoretical issue. Intuitively, one expects modest variations in parameters to have little effect on the predictors because they are interpolators. In practice, "Studentized" cross-validation residuals (leave-one-out predictions of the data normalized by standard error) have been successfully used to gauge the "legitimacy" of such usage (for examples and additional references see Schonlau and Welch [25] and Aslett et al. [1]). Only recently, Nagy, Loeppky and Welch [19] have reported simulations indicating reasonably close prediction accuracy of the plug-in MLE predictions to Bayes (Jeffreys priors) predictions in dimensions 1–10 when the number of computer runs $= 7 \times$ dimension.

*The posterior distributions.* Restricting to the test bed problem, we simplify the notation by referring only to $\boldsymbol{\delta}$, the deviation of $\mathbf{x}$ from the nominal inputs $\mathbf{x}_{\text{nom}}$, and rewrite (8) and (9), $\forall i \in I$, as

(12) $$w_i^R(\boldsymbol{\delta}^*) = w_i^M(\boldsymbol{\delta}^*, \mathbf{u}^*) + w_i^b,$$

$$w_{ir}^F(\boldsymbol{\delta}^*) = w_i^R(\boldsymbol{\delta}^*) + \varepsilon_{ir},$$

where the $\varepsilon_{ir}$ are independent $N(0, \sigma_i^2)$.

The field data can be summarized by the (independent) sufficient statistics

$$\bar{w}_i^F = \frac{1}{7}\sum_{r=1}^{7} w_{ir}^F(\boldsymbol{\delta}^*), \qquad s_i^2 = \sum_{r=1}^{7}(w_{ir}^F(\boldsymbol{\delta}^*) - \bar{w}_i^F)^2.$$

(We drop the argument $\boldsymbol{\delta}^*$ for these statistics, since the statistics are actual numbers given from the data.) Key facts to be retained, using (11) and (12) and properties of normal distributions are

$$\frac{s_i^2}{\sigma_i^2} \,\Big|\, \sigma_i^2 \sim \chi_6^2,$$

$$(13) \quad \bar{w}_i^F \mid w_i^M(\boldsymbol{\delta}^*, \mathbf{u}^*), w_i^b, \sigma_i^2 \sim N(w_i^M(\boldsymbol{\delta}^*, \mathbf{u}^*) + w_i^b, \tfrac{1}{7}\sigma_i^2),$$

$$\bar{w}_i^F \mid \boldsymbol{\delta}^*, \mathbf{u}^*, \mathbf{w}_i^M, w_i^b, \sigma_i^2 \sim N(\widehat{m}_i^M(\boldsymbol{\delta}^*, \mathbf{u}^*) + w_i^b, \widehat{V}_i^M(\boldsymbol{\delta}^*, \mathbf{u}^*) + \tfrac{1}{7}\sigma_i^2),$$

where (11) is used to get the last expression.

Let $\mathbf{w}^b$, $\boldsymbol{\tau}^2$, $\boldsymbol{\sigma}^2$, $\boldsymbol{\delta}^*$ and $\mathbf{u}^*$ denote the vectors of the indicated parameters and write their prior distribution as

$$\pi(\mathbf{w}^b, \boldsymbol{\tau}^2, \boldsymbol{\sigma}^2, \boldsymbol{\delta}^*, \mathbf{u}^*)$$

$$(14) \qquad = \pi(\mathbf{w}^b \mid \boldsymbol{\tau}^2) \times \pi(\boldsymbol{\delta}^*, \mathbf{u}^*, \boldsymbol{\tau}^2 \mid \boldsymbol{\sigma}^2) \times \pi(\boldsymbol{\sigma}^2)$$

$$= \prod_{i \in I} \pi(w_i^b \mid \tau_{j(i)}^2) \times \left[ \prod_{i=1}^7 \pi(\delta_i^*) \prod_{i=1}^2 \pi(u_i^*) \prod_{j=0}^{12} \pi(\tau_j^2 \mid \{\sigma_i^2\}) \right]$$

$$\times \prod_{i \in I} \pi(\sigma_i^2).$$

The data, from field and computer model runs, can be summarized as $\mathbf{D} = \{\bar{w}_i^F, s_i^2, \mathbf{w}_i^M; i = 1, \ldots, 289\}$. Using (11), (13), (10) and (14), together with standard computations involving normal distributions, it is straightforward to obtain the posterior distribution of all unknowns as

$$\pi_{post}(w^M(\boldsymbol{\delta}^*, \mathbf{u}^*), \mathbf{w}^b, \boldsymbol{\delta}^*, \mathbf{u}^*, \boldsymbol{\sigma}^2, \boldsymbol{\tau}^2 \mid \mathbf{D})$$

$$= \pi_{post}(w^M(\boldsymbol{\delta}^*, \mathbf{u}^*) \mid \mathbf{w}^b, \boldsymbol{\delta}^*, \mathbf{u}^*, \boldsymbol{\sigma}^2, \mathbf{D})$$

$$(15) \qquad \times \pi_{post}(\mathbf{w}^b \mid \boldsymbol{\delta}^*, \mathbf{u}^*, \boldsymbol{\sigma}^2, \boldsymbol{\tau}^2, \mathbf{D})$$

$$\times \pi_{post}(\boldsymbol{\delta}^*, \mathbf{u}^*, \boldsymbol{\tau}^2 \mid \boldsymbol{\sigma}^2, \mathbf{D})$$

$$\times \pi_{post}(\boldsymbol{\sigma}^2 \mid \mathbf{D}),$$

where

$$\pi_{post}(w^M(\boldsymbol{\delta}^*, \mathbf{u}^*) \mid \mathbf{w}^b, \boldsymbol{\delta}^*, \mathbf{u}^*, \boldsymbol{\sigma}^2, \mathbf{D}) \sim \prod_{i \in I} N(m_{1i}, V_{1i});$$

$$m_{1i} = \frac{\widehat{V}_i^M(\boldsymbol{\delta}^*, \mathbf{u}^*)}{\widehat{V}_i^M(\boldsymbol{\delta}^*, \mathbf{u}^*) + (1/7)\sigma_i^2}(\bar{w}_i^F - w_i^b)$$

$$(16) \qquad + \frac{(1/7)\sigma_i^2}{\widehat{V}_i^M(\boldsymbol{\delta}^*, \mathbf{u}^*) + (1/7)\sigma_i^2}(\widehat{m}_i^M(\boldsymbol{\delta}^*, \mathbf{u}^*)),$$

$$V_{1i} = \frac{\widehat{V}_i^M(\boldsymbol{\delta}^*, \mathbf{u}^*)(1/7)\sigma_i^2}{\widehat{V}_i^M(\boldsymbol{\delta}^*, \mathbf{u}^*) + (1/7)\sigma_i^2}$$

and

$$\pi_{post}(\mathbf{w}^b \mid \boldsymbol{\delta}^*, \mathbf{u}^*, \boldsymbol{\sigma}^2, \boldsymbol{\tau}^2, \mathbf{D}) \sim \prod_{i \in I} N(m_{2i}, V_{2i});$$

(17)
$$m_{2i} = \frac{\tau_{j(i)}^2}{\widehat{V}_i^M(\boldsymbol{\delta}^*, \mathbf{u}^*) + (1/7)\sigma_i^2 + \tau_{j(i)}^2} (\bar{w}_i^F - \widehat{m}_i^M(\boldsymbol{\delta}^*, \mathbf{u}^*)),$$

$$V_{2i} = \frac{\tau_{j(i)}^2(\widehat{V}_i^M(\boldsymbol{\delta}^*, \mathbf{u}^*) + (1/7)\sigma_i^2)}{\widehat{V}_i^M(\boldsymbol{\delta}^*, \mathbf{u}^*) + (1/7)\sigma_i^2 + \tau_{j(i)}^2}.$$

The third factor in (15) is

(18)
$$\pi_{post}(\boldsymbol{\delta}^*, \mathbf{u}^*, \boldsymbol{\tau}^2 \mid \boldsymbol{\sigma}^2, \mathbf{D})$$
$$\propto L(\overline{\mathbf{w}}^F, \mathbf{s}^2 \mid \boldsymbol{\delta}^*, \mathbf{u}^*, \boldsymbol{\sigma}^2, \boldsymbol{\tau}^2) \times \pi(\boldsymbol{\delta}^*, \mathbf{u}^*, \boldsymbol{\tau}^2 \mid \boldsymbol{\sigma}^2),$$

where the marginal likelihood $L$ found by integrating out $\mathbf{w}^b$ and $w^M(\boldsymbol{\delta}^*, \mathbf{u}^*)$ in the product of the full likelihood and $\pi(\mathbf{w}^b \mid \boldsymbol{\tau}^2)$, is

$$L(\overline{\mathbf{w}}^F, \mathbf{s}^2 \mid \boldsymbol{\delta}^*, \mathbf{u}^*, \boldsymbol{\sigma}^2, \boldsymbol{\tau}^2) = \prod_{i \in I} \frac{1}{\sqrt{\widehat{V}_i^M(\boldsymbol{\delta}^*, \mathbf{u}^*) + (1/7)\sigma_i^2 + \tau_{j(i)}^2}}$$
$$\times \exp\left\{ -\frac{1}{2}\left( \frac{(\bar{w}_i^F - \widehat{m}_i^M(\boldsymbol{\delta}^*, \mathbf{u}^*))^2}{\widehat{V}_i^M(\boldsymbol{\delta}^*, \mathbf{u}^*) + (1/7)\sigma_i^2 + \tau_{j(i)}^2} \right) \right\}.$$

Finally, the fourth factor in (15) is

(19)
$$\pi_{post}(\boldsymbol{\sigma}^2 \mid \mathbf{D}) \propto \left[ \prod_{i \in I} \frac{1}{(\sigma_i^2)^3} \exp\left\{ -\frac{s_i^2}{2\sigma_i^2} \right\} \right]$$
$$\times \int L(\overline{\mathbf{w}}^F, \mathbf{s}^2 \mid \boldsymbol{\delta}^*, \mathbf{u}^*, \boldsymbol{\sigma}^2, \boldsymbol{\tau}^2) \, d\boldsymbol{\delta}^* \, d\mathbf{u}^* \, d\boldsymbol{\tau}^2.$$

At this point we make an approximation, and ignore the integral in (19); that is, we simply utilize the replicate observations to determine the posteriors for the $\sigma_i^2$. The reason for this is not computational; indeed, one can include the $\sigma_i^2$ in the posterior in (18) and deal with them by a Metropolis algorithm. Instead, the motivation is what we call *modularization*, which is meant to indicate that it can be better to separately analyze pieces of the problem than to perform one global Bayesian analysis. The difficulty here is that there is a significant confounding in the posterior distribution between the calibration parameters, the bias function, and the $\sigma_i^2$ and this, for instance, allows bias to be replaced by larger $\sigma_i^2$. Here we have seven replicate observations for each $\sigma_i^2$, so simply utilizing the replicate observation posteriors and preventing the confounding has intuitive appeal. (Formalizing this argument is not so easy. The difficulty occurs because part of the model—the modeling of the bias—is quite uncertain. Better or more robust modeling of the bias may correct the problem within a full Bayesian analysis, but the difficulty of doing so argues for the simpler modular approach. We will discuss these issues more fully elsewhere.)

Simulating from (16) and (17) and the first factor of (19) is, of course, trivial, but simulating from (18) requires MCMC methodology. Given the complexity of the problem, the MCMC requires careful choice of proposal distributions in order to achieve suitable mixing. Discussion of these proposals is relegated to Appendix B because of the level of detail needed to describe them, but we note that these are technically crucial for the methodology to work and required extensive exploration.

The end result of the simulation is a sample of draws from the posterior distribution in (15): each saved draw from the first factor of (19) is used to generate the MCMC sample for the third factor, with both being used to generate a draw from the second and the first factors, using (16) and (17). We saved every 200th draw from 200,000 MCMC iterations for the third factor, thereby obtaining a final sample of 1000 draws,

$$(20) \qquad \{w^{M,h}(\boldsymbol{\delta}^{*h}, \mathbf{u}^{*h}), \mathbf{w}^{bh}, \boldsymbol{\delta}^{*h}, \mathbf{u}^{*h}, \boldsymbol{\sigma}^{2h}, \boldsymbol{\tau}^{2h}; h = 1, \ldots, 1000\}.$$

The results below are based on this sample from the posterior.

## 5. Results.

5.1. *Estimates of $\boldsymbol{\delta}^*, \mathbf{u}^*$.* Histograms for $\boldsymbol{\delta}^*, \mathbf{u}^*$ (Figure 4) are obtained by forming a histogram for each component of $\boldsymbol{\delta}^*, \mathbf{u}^*$ from the corresponding elements in (20). The calibration parameters are moderately affected by the data but, of the input variables, only $x_5$ and $x_6$ have posteriors that are significantly different from the priors. The posterior for $x_5$ is piled up at the end of the allowed range for the variable, which suggests the (undesirable) possibility that this uncertain input is being used as a tuning parameter to better fit the model; a case could be made for preventing this by additional modularization.

5.2. *Estimation of bias and reality.* Posterior distributions of the bias and reality curves are obtained by recombining the wavelets with the posterior wavelet coefficients from (20). For instance, the posterior distribution of $b$ is represented by the sample curves

$$(21) \qquad b^h(t) = \sum_{i \in I} w_i^{bh} \Psi_i(t), \qquad h = 1, \ldots, 1000.$$

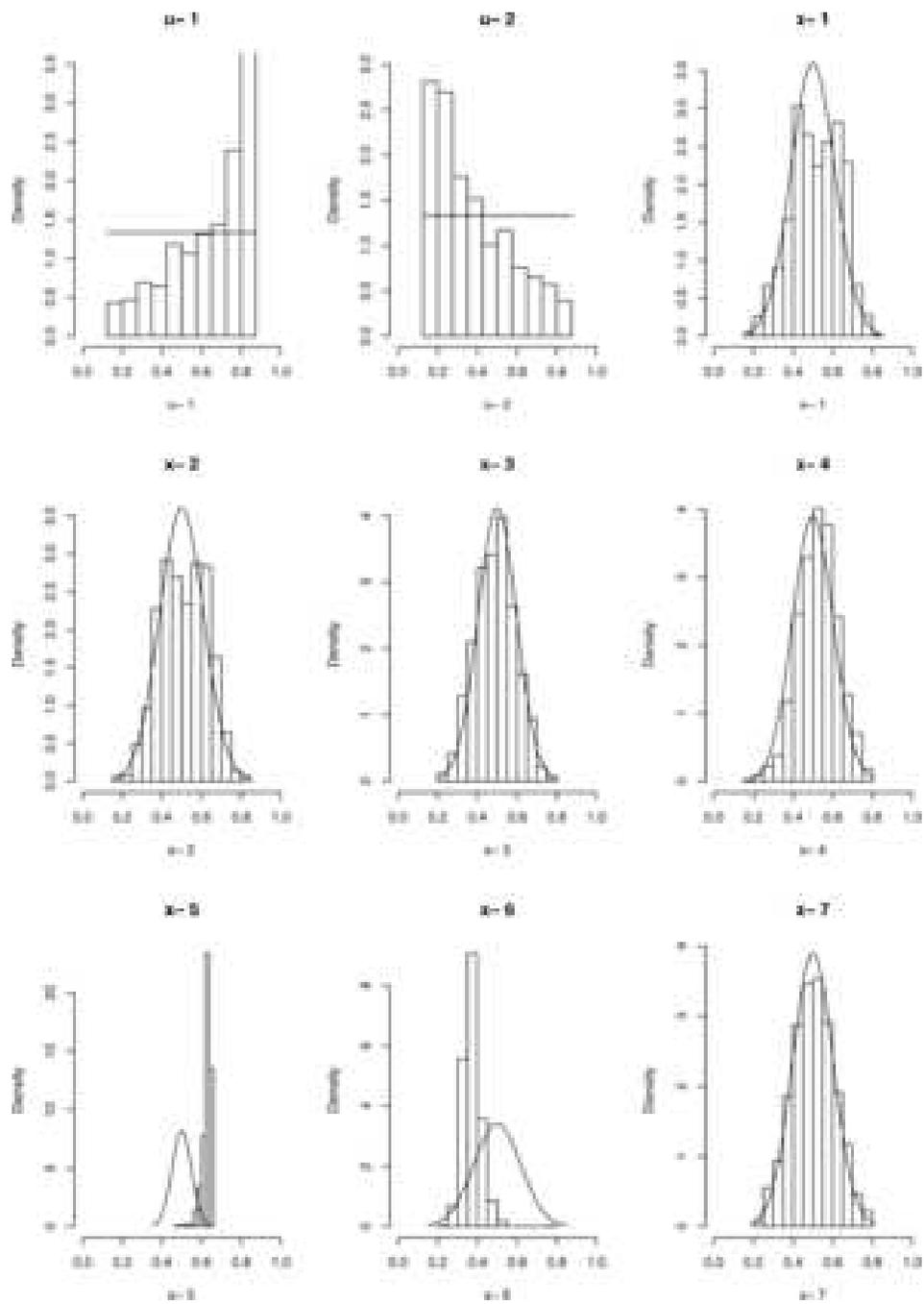

FIG. 4.   *Histogram of the posterior draws for the two calibration parameters ($u_1$ and $u_2$), and the seven input parameters ($x_1$ through $x_7$), along with their priors (solid lines).*

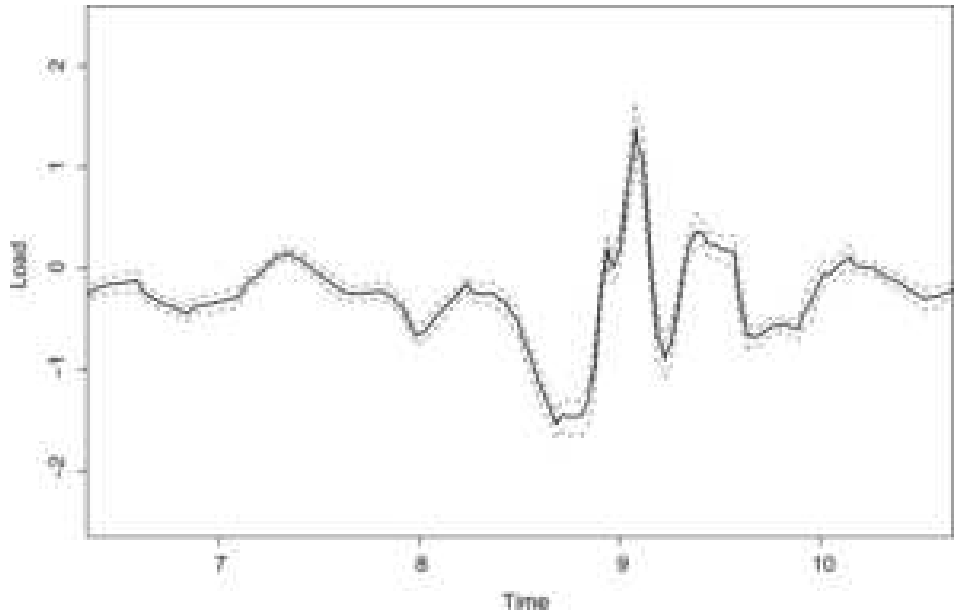

FIG. 5. *The estimate of the bias function (solid line) with 90% tolerance bounds (dashed lines) for suspension Site 1 and at Region 1.*

The posterior mean curve, $\widehat{b}(t) = \frac{1}{1000}\sum_{h=1}^{1000} b^h(t)$, is plotted as the solid line in Figure 5. The uncertainty of this estimate of $b$ is quantified by producing upper and lower uncertainty (tolerance) bounds at each $t$ by, for example, taking the lower $\alpha/2$ and upper $1 - \alpha/2$ quantiles of the posterior distribution of $b(t)$, that is,

$$L^b(t) = \frac{\alpha}{2} \text{ quantile of } \{b^h(t); h = 1, \ldots, 1000\},$$

(22)

$$U^b(t) = \left(1 - \frac{\alpha}{2}\right) \text{ quantile of } \{b^h(t); h = 1, \ldots, 1000\}.$$

These lower and upper bounds are also plotted in Figure 5. It is apparent in Figure 5 that the bias function is significantly different from 0, especially in the neighborhood of 8.7 and 9.1.

The bounds in (22) are symmetrically defined. Alternative tolerance bounds can be defined by only requiring that $100\alpha\%$ of the curves lie outside the bounds; a useful choice would satisfy this condition and minimize the height of the bounds, $U^b(t) - L^b(t)$.

Figures 4 and 5 provide marginal distributions of $\mathbf{u}^*$, $\mathbf{x}_{\text{nom}} + \boldsymbol{\delta}^*$ and the bias, but it is important to note that these are highly dependent in the posterior. Hence most analyses involving these quantities must be based on their joint, rather than marginal, distributions.

Estimating reality with uncertainty bounds is done similarly: take the sample of wavelet coefficients $w_i^{Rh} = w_i^{Mh}(\boldsymbol{\delta}^{*h}, \mathbf{u}^{*h}) + w_i^{bh}$ and form

$$
\begin{aligned}
y^{Rh}(t) &= \sum_i w_i^{Rh} \Psi_i(t), \\
\widehat{y}^R(t) &= \frac{1}{1000} \sum_h y^{Rh}(t), \\
L^R(t) &= \frac{\alpha}{2} \text{ quantile of } \{y^{Rh}(t); h = 1, \dots, 1000\}, \\
U^R(t) &= \left(1 - \frac{\alpha}{2}\right) \text{ quantile of } \{y^{Rh}(t); h = 1, \dots, 1000\}.
\end{aligned}
$$

(23)

We call $\widehat{y}^R(t)$ the *bias-corrected prediction of reality.* Figure 6 exhibits the bias-corrected prediction and associated uncertainty band.

Figure 6 further shows a comparison between bias-corrected prediction and pure model prediction, the latter being defined as

(24) $$\widehat{y}^M(t) = \sum_i \widehat{m}_i^M(\widehat{\boldsymbol{\delta}}, \widehat{\mathbf{u}}) \Psi_i(t),$$

where $\widehat{\boldsymbol{\delta}} = \frac{1}{1000} \sum_h \boldsymbol{\delta}^{*h}$ and $\widehat{\mathbf{u}} = \frac{1}{1000} \sum_h \mathbf{u}^{*h}$ and $\widehat{m}_i^M(\widehat{\boldsymbol{\delta}}, \widehat{\mathbf{u}})$ is the posterior mean of the wavelet coefficients with plugged-in estimates for the unknown parameters [use (11)].

In practice, it may be that running the computer model after estimating $\boldsymbol{\delta}^*, \mathbf{u}^*$ is feasible. Then an alternative (and preferred) pure model prediction is $y^M(\widehat{\boldsymbol{\delta}}, \widehat{\mathbf{u}}; t)$.

Assessing the uncertainty for predicting reality by the pure model prediction [see (24)] can be done by considering samples $\{y^{Rh}(t) - \widehat{y}^M(t)\}$ and forming bounds. If a new model run producing $y^M(\widehat{\boldsymbol{\delta}}, \widehat{\mathbf{u}}; t)$ is used instead, then the computation should be redone with this additional model run included, but a typically accurate approximation is to simply consider $\{y^{Rh}(t) - y^M(\widehat{\boldsymbol{\delta}}, \widehat{\mathbf{u}}; t)\}$ and proceed as before. Here it may be useful to consider asymmetric bounds because the pure model predictions may lie entirely above or below the realizations of reality. But plots like that of Figure 6 already show the gap between pure model prediction and reality.

### 5.3. *Predicting a new run; same system, same vehicle (new run).* In some prediction settings, not necessarily the one of the test bed, there is interest in predicting a new *field run* with the same inputs (and the same system). Prediction is done by adding in draws $\varepsilon_i^h$ from a $N(0, \sigma_i^{2h})$ distribution and then following the same prescription as in (23) to form $\widehat{y}^F(t)$ and corresponding uncertainty bounds. This, of course, produces wider uncertainty bounds.

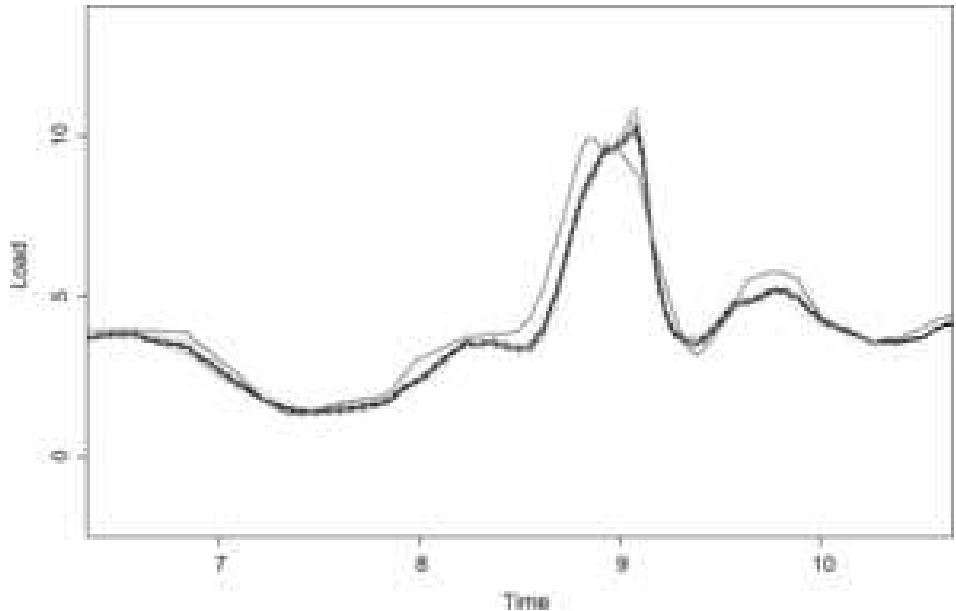

Fig. 6. *Bias-corrected prediction of reality (solid black line) with 90% tolerance bands (dashed black lines), pure model prediction (solid grey line) and field runs (light grey lines).*

5.4. *Extrapolating.* There are many follow-on settings where prediction is called for. We single out three such: (i) the same vehicle except that $\mathbf{x}_{\text{nom}}$ changes; (ii) same vehicle type but new components, that is, the same nominal values and prior distributions for the manufacturing variations but a new realization of the $\boldsymbol{\delta}$'s from the prior distribution; (iii) a new vehicle, different nominal $(x_1, \ldots, x_7)$ inputs, but the same prior distributions for the $\boldsymbol{\delta}$'s.

Any analysis we perform for the test bed is severely constrained by the fact that we have limited field data on just one set of inputs measured with error.

5.4.1. *Same vehicle, different load.* Here the same vehicle was tested but with added mass. This causes a known change in $\mathbf{x}_{\text{nom}}$ with everything else (including the $\boldsymbol{\delta}$'s) remaining unchanged. The effect of a simple change in inputs such as this can be addressed using the difference in computer model runs at the *nominal* values of the inputs. More formally, suppose we add a computer model run at $(\mathbf{x}_{\text{nom}} + \boldsymbol{\Delta}, \mathbf{u}_{\text{nom}})$ where $\boldsymbol{\Delta}$ is the modest change in the nominal inputs ($\boldsymbol{\Delta}$ has only one nonzero coordinate if the only change is in the mass). Suppose we also have a run at the old nominals $(\mathbf{x}_{\text{nom}}, \mathbf{u}_{\text{nom}})$; if not, use the GASP prediction based on the existing runs. Our assumption then is that (with $\mathbf{x}^*$ referring to the true unknown input values for the

original system)

$$y^M(\mathbf{x}^* + \boldsymbol{\Delta}, \mathbf{u}^*; t) - y^M(\mathbf{x}^*, \mathbf{u}^*; t)$$
$$\simeq y^M(\mathbf{x}_{\mathrm{nom}} + \boldsymbol{\Delta}, \mathbf{u}_{\mathrm{nom}}; t) - y^M(\mathbf{x}_{\mathrm{nom}}, \mathbf{u}_{\mathrm{nom}}; t) \equiv \mathbf{D}(t).$$

We can then make predictions under the new inputs by simply adding $\mathbf{D}(t)$ to the old predictions.

This is illustrated in Figure 7; the given bias corrected prediction and tolerance bands for the vehicle with the additional mass are simply the results of Section 5.3 translated by $D(t)$. The grey solid line is the pure model prediction. The light grey line is the actual result from a field test of the system with added mass, and the strategy appears to have successfully extrapolated in the critical Region 1. Again, the pure model prediction is not successful.

5.4.2. *New vehicle of the same type.* In this setting the nominal values $\mathbf{x}_{\mathrm{nom}}$ remain the same but the new $\boldsymbol{\delta}$'s are random draws from their *prior* (population) distribution and are, therefore, different from those for the field-tested system. This is of particular interest in practice, in that it is prediction for the population of vehicles of the given type that is of prime interest, rather than just prediction for the single system/vehicle tested.

The calibration parameters $\mathbf{u}^*$ do not change; they belong to the model and, if physically real, are inherently the same for all systems/vehicles of the same type. Denote the parameters of the new components by $\mathbf{z}_{\mathrm{new}} = (\mathbf{x}_{\mathrm{nom}} + \boldsymbol{\delta}_{\mathrm{new}}, \mathbf{u}^*)$. The input parameters of the computer runs remain $\mathbf{z}_k = (\mathbf{x}_{\mathrm{nom}} + \boldsymbol{\delta}_k, \mathbf{u}_k)$; and $\mathbf{z}^* = (\mathbf{x}_{\mathrm{nom}} + \boldsymbol{\delta}^*, \mathbf{u}^*)$ are the true values for the tested system. Denote the associated model wavelet coefficients for the new components by $w^M(\mathbf{z}_{\mathrm{new}})$.

Since $\boldsymbol{\delta}_{\mathrm{new}}$ is independent of $(\mathbf{w}^b, \boldsymbol{\delta}^*, \mathbf{u}^*, \boldsymbol{\sigma}^2, \boldsymbol{\tau}^2, \mathbf{D})$, the predictive (posterior) distribution is (with the $L_i$'s denoting likelihood terms arising from the data)

$$
\begin{aligned}
&\pi_{post}(w^M(\mathbf{z}_{\mathrm{new}}), w^M(\mathbf{z}^*), \mathbf{w}^b, \boldsymbol{\delta}_{\mathrm{new}}, \boldsymbol{\delta}^*, \mathbf{u}^*, \boldsymbol{\sigma}^2, \boldsymbol{\tau}^2 \mid \mathbf{D}) \\
&\quad \propto \pi(\boldsymbol{\delta}_{\mathrm{new}}) \times \pi(\mathbf{w}^b, \boldsymbol{\delta}^*, \mathbf{u}^*, \boldsymbol{\sigma}^2, \boldsymbol{\tau}^2) \\
&\qquad \times L_1(w^M(\mathbf{z}_{\mathrm{new}}), w^M(\mathbf{z}^*), \{w_i^M(z_k)\} \mid \boldsymbol{\delta}_{\mathrm{new}}, \boldsymbol{\delta}^*, \mathbf{u}^*) \\
&\qquad \times L_2(\{\bar{w}_i\}, \{s_i^2\} \mid w^M(\mathbf{z}^*), \mathbf{w}^b, \boldsymbol{\sigma}^2) \\
&\quad \propto \pi_{post}(w^M(\mathbf{z}_{\mathrm{new}}) \mid w^M(\mathbf{z}^*), \{w_i^M(z_k)\}, \boldsymbol{\delta}_{\mathrm{new}}, \boldsymbol{\delta}^*, \mathbf{u}^*) \\
&\qquad \times \pi(\boldsymbol{\delta}_{\mathrm{new}}) \times \pi(\mathbf{w}^b, \boldsymbol{\delta}^*, \mathbf{u}^*, \boldsymbol{\sigma}^2, \boldsymbol{\tau}^2) \\
&\qquad \times L_3(w^M(\mathbf{z}^*), \{w_i^M(z_k)\} \mid \boldsymbol{\delta}^*, \mathbf{u}^*) \\
&\qquad \times L_2(\{\bar{w}_i\}, \{s_i^2\} \mid w^M(\mathbf{z}^*), \mathbf{w}^b, \boldsymbol{\sigma}^2).
\end{aligned}
\tag{25}
$$

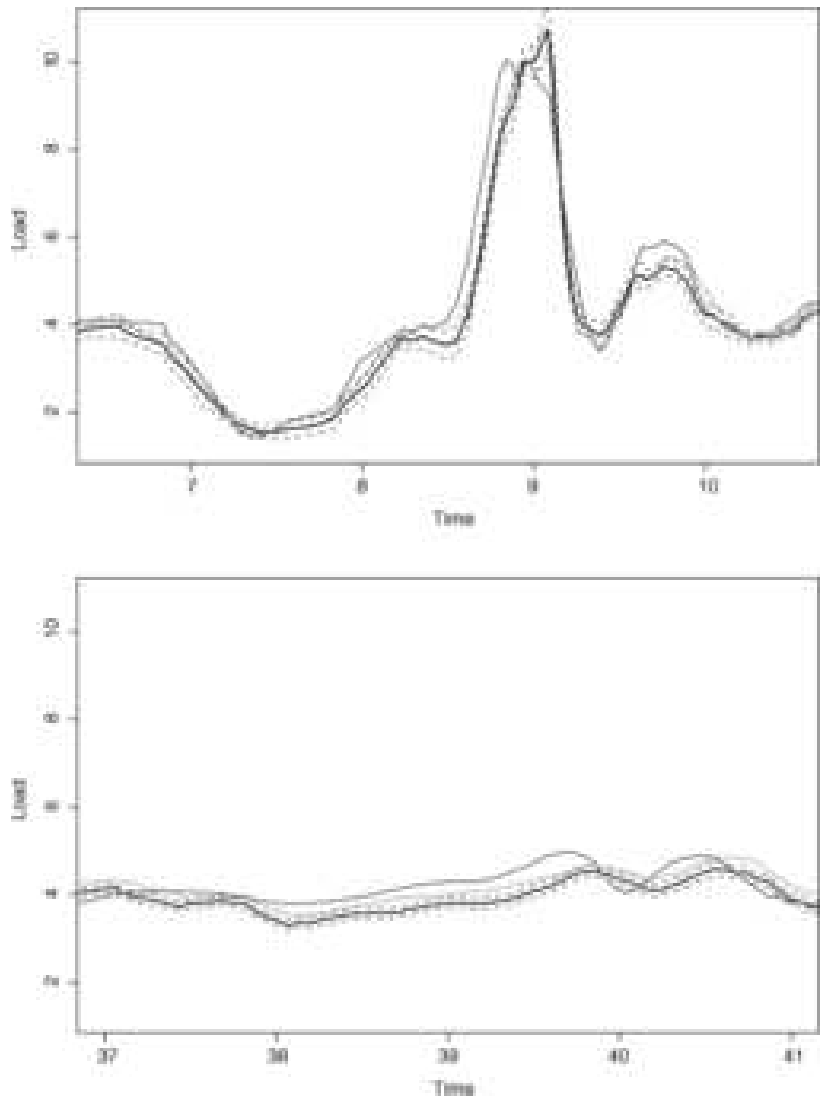

FIG. 7. *Prediction at Site 1 for both Regions 1 (top) and 2 (bottom) for a run with additional mass. The solid black line is bias-corrected prediction, the dashed black lines are 90% tolerance bands, the grey line is pure model prediction and the field run is the light grey line.*

To sample from (25), note that the last three factors in the expression yield exactly the same posterior for $(w^M(\mathbf{z}^*), \mathbf{w}^b, \boldsymbol{\delta}^*, \mathbf{u}^*, \boldsymbol{\sigma}^2, \boldsymbol{\tau}^2)$ as before (with the same modularization used), so the draws from the previous MCMC can be used in the new computations. Since $\boldsymbol{\delta}_{\text{new}}$ can be drawn from its prior $\pi(\boldsymbol{\delta}_{\text{new}})$, it only remains to draw from $\pi_{post}(w^M(\mathbf{z}) \mid w^M(\mathbf{z}^*), \{w_i^M(z_k)\}, \boldsymbol{\delta}_{\text{new}},$ $\boldsymbol{\delta}^*, \mathbf{u}^*)$. But this is simply the GASP distribution where $w^M(\mathbf{z}^*)$ has to be added to the model run data. Therefore, one simply determines the GASP for the augmented runs $\mathbf{w}_i^{M0} = (w_i^M(\mathbf{z}_1), w_i^M(\mathbf{z}_2), \ldots, w_i^M(\mathbf{z}_k), w_i^M(\mathbf{z}^*))$, that is,

$$(26) \qquad w_i^M(\mathbf{z}_{\text{new}}) \mid \mathbf{w}_i^{M0}, \widehat{\boldsymbol{\theta}}_i^M \sim N(\widehat{m}_i^{M0}(\mathbf{z}_{\text{new}}), \widehat{V}_i^{M0}(\mathbf{z}_{\text{new}})),$$

with $\widehat{\boldsymbol{\theta}}_i^M$ as in Section 4 and

$$\widehat{m}_i^{M0}(\mathbf{z}_{\text{new}}) = \widehat{\mu}_i^M + \widehat{\gamma}_i^{M0}(\mathbf{z}_{\text{new}})'(\widehat{\boldsymbol{\Gamma}}_i^{M0})^{-1}(\mathbf{w}_i^{M0} - \widehat{\mu}_i^M \mathbf{1}),$$

$$\widehat{V}_i^{M0}(\mathbf{z}_{\text{new}}) = \frac{1}{\widehat{\lambda}_i^M} - \widehat{\gamma}_i^{M0}(\mathbf{z}_{\text{new}})'(\widehat{\boldsymbol{\Gamma}}_i^{M0})^{-1}\widehat{\gamma}_i^{M0}(\mathbf{z}_{\text{new}}),$$

where $\widehat{\gamma}_i^{M0}(\mathbf{z}_{\text{new}}) = (1/\widehat{\lambda}_i^M)(\widehat{c}_i^M(\mathbf{z}_1, \mathbf{z}_{\text{new}}), \ldots, \widehat{c}_i^M(\mathbf{z}_k, \mathbf{z}_{\text{new}}), \widehat{c}_i^M(\mathbf{z}^*, \mathbf{z}_{\text{new}}))'$ and $\widehat{\boldsymbol{\Gamma}}_i^{M0}$ is obtained by appending the column $\widehat{\gamma}_i^{M0}(\mathbf{z}^*)$ and the row $\widehat{\gamma}_i^{M0}(\mathbf{z}^*)'$ to $\widehat{\boldsymbol{\Gamma}}_i^M$.

Application of these expressions yields draws $h = 1, \ldots, 1000$ from the posterior distribution of the $i$th wavelet coefficient for the new system as

$$w_i^{Fh}(\mathbf{z}^h) = w_i^M(\mathbf{z}^h) + w_i^{bh} + \varepsilon_i^h,$$

where $\varepsilon_i^h \sim N(0, \sigma_i^{2h})$. Figure 8 plots the predictions of the new system with uncertainty bands. The uncertainty has increased as compared with Figure 6 not only because the prior for $\boldsymbol{\delta}_{\text{new}}$ is used rather than the posterior for the tested vehicle, but also because these are predictions of field runs (instead of reality).

5.4.3. *New vehicle with new nominals.* The primary engineering use of computer models is to extrapolate to a system with new nominals when there are no new field data. This will require strong assumptions, especially about the bias. The simplest assumption about the bias, which we make here, is that the new system has the same bias function as the old system. The calibration parameters $\mathbf{u}^*$ are also assumed to remain the same. We use the joint—and highly dependent—posterior distribution of the bias and $\mathbf{u}^*$ from the original system extensively in what follows.

The new system has the same I/U map as the original system, but with new nominal values $\mathbf{x}_{\text{nom}}^B$ (which we refer to as Condition B). The new $\boldsymbol{\delta}$'s are taken to have the same priors as for Condition A. The same 65-point design on $(\boldsymbol{\delta}, \mathbf{u})$ was used as before with the addition of one central value. Again one run failed, leaving 65 usable model runs. Registration of the output curves is unnecessary because there are no field data and the computer runs are assumed to be inherently registered. The resulting computer runs were passed through the same wavelet decomposition as before, retaining only

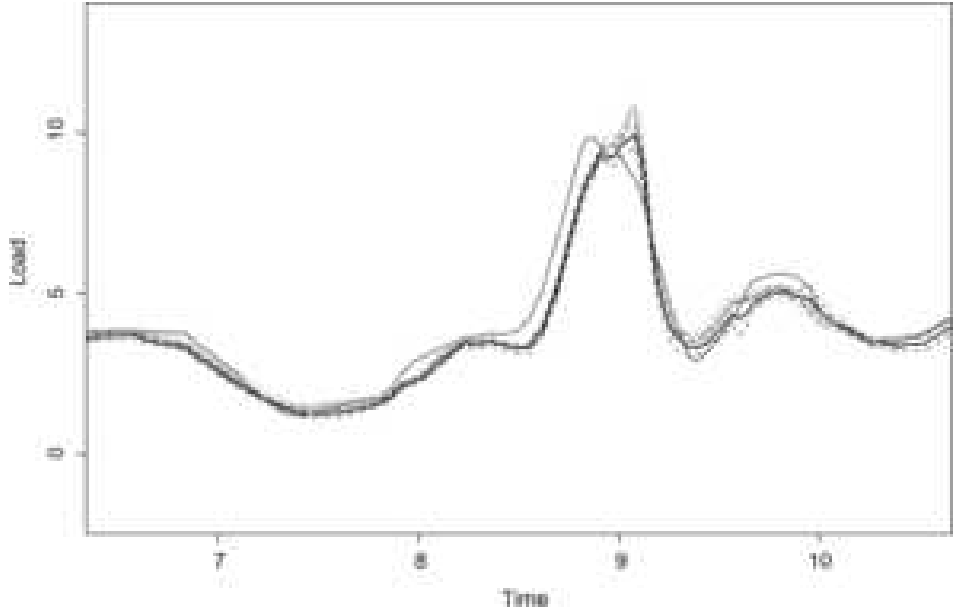

Fig. 8. *Predictions for a new vehicle of the same type. The solid black line is bias-corrected prediction, the dashed black lines are 90% tolerance bands, the grey line is pure model prediction and the light grey lines are field runs.*

those coefficients that appeared earlier. The resulting GASP for the $i$th wavelet coefficient is

$$(27) \qquad w_i^{BM}(\mathbf{z}) \mid \mathbf{w}_i^{BM}, \widehat{\boldsymbol{\theta}}_i^{BM} \sim N(\widehat{m}_i^{BM}(\mathbf{z}), \widehat{V}_i^{BM}(\mathbf{z})),$$

exactly as in (11).

This new GASP analysis is done only with Condition B data. A GASP analysis combining the model runs for Condition B *and* the model runs for the original system is not used because the changes in the nominals are too large to safely assume connections between the computer model runs for the two systems.

The situation now is analogous to that of the previous argument for new vehicles of the same type with the same nominals. In the current case, again using the independence of $\boldsymbol{\delta}^B$ from the other unknowns, the predictive (posterior) distribution of the relevant unknowns can be written as

$$\pi(\mathbf{w}^{BM}(\mathbf{z}^B), \mathbf{w}^b, \boldsymbol{\delta}^B, \mathbf{u}^*, \boldsymbol{\sigma}^2, \boldsymbol{\tau}^2 \mid \mathbf{D})$$
$$= \pi(\mathbf{w}^{BM}(\mathbf{z}^B) \mid \mathbf{w}^b, \boldsymbol{\delta}^B, \mathbf{u}^*, \boldsymbol{\sigma}^2, \boldsymbol{\tau}^2, \mathbf{D})$$
$$\times \pi(\boldsymbol{\delta}^B) \times \pi(\mathbf{w}^b, \mathbf{u}^*, \boldsymbol{\sigma}^2, \boldsymbol{\tau}^2 \mid \mathbf{D})$$
$$= \pi(\mathbf{w}^{BM}(\mathbf{z}^B) \mid \boldsymbol{\delta}^B, \mathbf{u}^*, \mathbf{D})$$
$$\times \pi(\boldsymbol{\delta}^B) \times \pi(\mathbf{w}^b, \mathbf{u}^*, \boldsymbol{\sigma}^2, \boldsymbol{\tau}^2 \mid \mathbf{D});$$

here $\pi(\mathbf{w}^{BM}(\mathbf{z}^B) \mid \boldsymbol{\delta}^B, \mathbf{u}^*, \mathbf{D})$ is just the GASP distribution in (27) and $\pi(\boldsymbol{\delta}^B)$ is the prior for $B$ inputs from the I/U map. Draws of $\mathbf{w}^b, \mathbf{u}^*, \boldsymbol{\sigma}^2, \boldsymbol{\tau}^2$ are made from the old posterior distribution for the original system. Because $\mathbf{w}^b$ and $\mathbf{u}^*$ are highly dependent in the posterior, they must be jointly sampled for the extrapolation; naive approaches–such as simply trying to add the bias from Figure 5 to the pure model prediction–will not succeed.

The "carry-over" assumptions for the bias and the field variances lead to draws from the posterior distribution of the wavelet coefficients for $B$ to be

$$w_i^{BFh}(\mathbf{z}^{Bh}) = w_i^{BM}(\mathbf{z}^h) + w_i^{bh} + \varepsilon_i^h,$$

where $\varepsilon_i^h \sim N(0, \sigma_i^{2h})$.

In the top of Figure 9 the prediction for $B$ is presented. Actual field data (8 replicate runs for $B$) were available afterward (not used in constructing the predictions and tolerance bands) and they are superimposed (in light grey) on the plots in the top of Figure 9; darker grey curves represent model runs. The effectiveness of carrying over the assumptions from $A$ to $B$ is apparent.

If a strong assumption, such as the constancy of bias, is to be made it is best to be extremely careful about implementing the assumption. Here, for instance, physics considerations might suggest that an assumption of constant multiplicative bias is more sensible than an assumption of constant additive bias. (Use of a multiplicative bias is mentioned in Fuentes, Guttorp and Challenor [10].) A standard way of implementing multiplicative bias is to repeat the above analysis using the log of the various curves. A simpler alternative is to transform the additive biases obtained above into multiplicative biases, and apply these multiplicative biases to the GASP draws under Condition B. Bias in the additive system can be written

$$b^h(t) = y^{Rh}(t) - y^{Mh}(t);$$

the corresponding multiplicative representation of the bias is

$$b_{\text{mult}}^h(t) = \frac{y^{Rh}(t)}{y^{Mh}(t)} - 1,$$

which would lead to draws from the posterior for reality under Condition B of

$$y^{BRh}(t) = y^{BMh}(t) \times (1 + b_{\text{mult}}^h(t)).$$

The bottom panel of Figure 9 presents the analogue of the top using the multiplicative bias. The additive and multiplicative predictions are not much different. The next section discusses another site in the suspension system for which the analysis in the paper was implemented, called Site 2. For this site, the multiplicative predictions under Condition B were noticeably better than the additive predictions, as indicated in Figure 10.

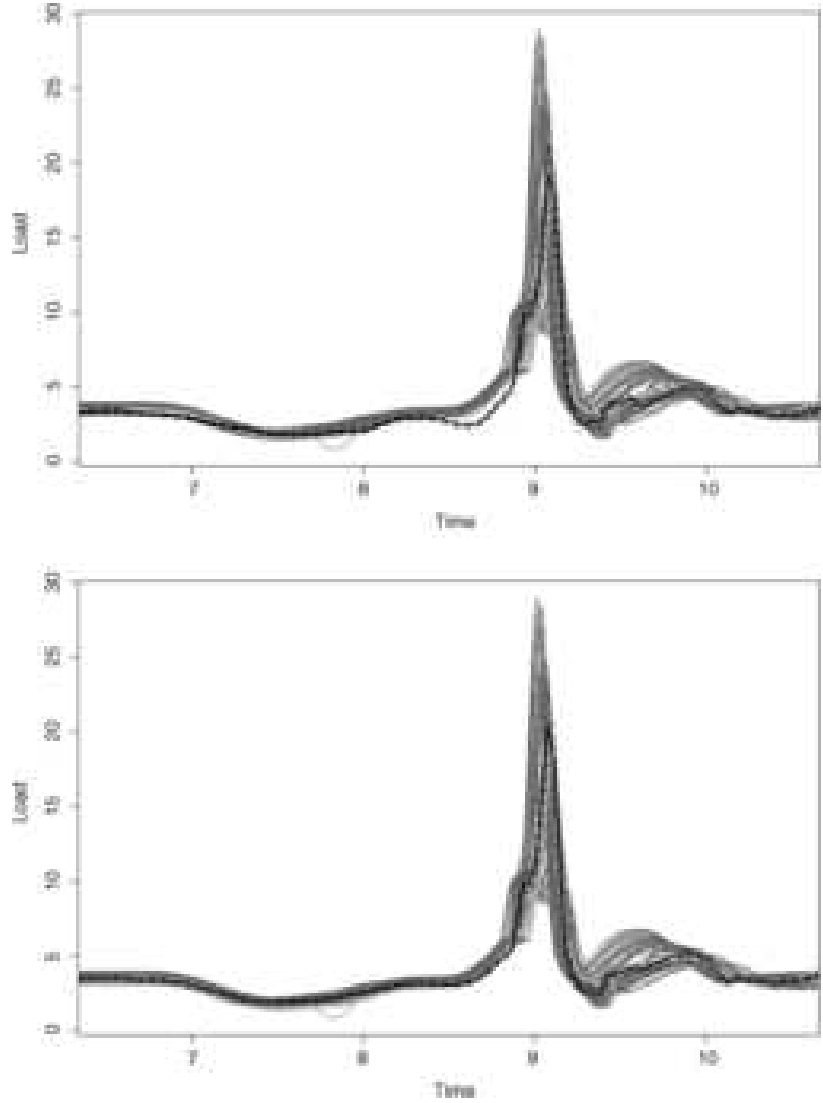

Fig. 9. *Prediction at Site 1 of a new vehicle under Condition* B *in Region 1. Top: additive bias. Bottom: multiplicative bias. The solid black line is bias-corrected prediction, the dashed black lines are 90% tolerance bands and the grey lines are model runs; the light grey lines are the field runs later provided and not used in the analysis.*

**6. Site 2 analyses.** Analyses for Site 2 of the system proceed in exactly the same way as those for Site 1. The posterior distributions for the calibration parameters $(u_1, u_2)$ as well as for $\boldsymbol{\delta}$ are in Figure 11. These are somewhat different than those for Site 1 in Figure 4. These parameters are, of course, the same for either site, but the limited data available at each site lead to somewhat different posterior distributions. In particular, $x_5$ no longer seems to be used as a tuning parameter, but it is possible that $u_2$ is being so used. A natural solution to help reduce such potential overtuning is to do a bivariate functional analysis of the two sites jointly. This is being pursued separately.

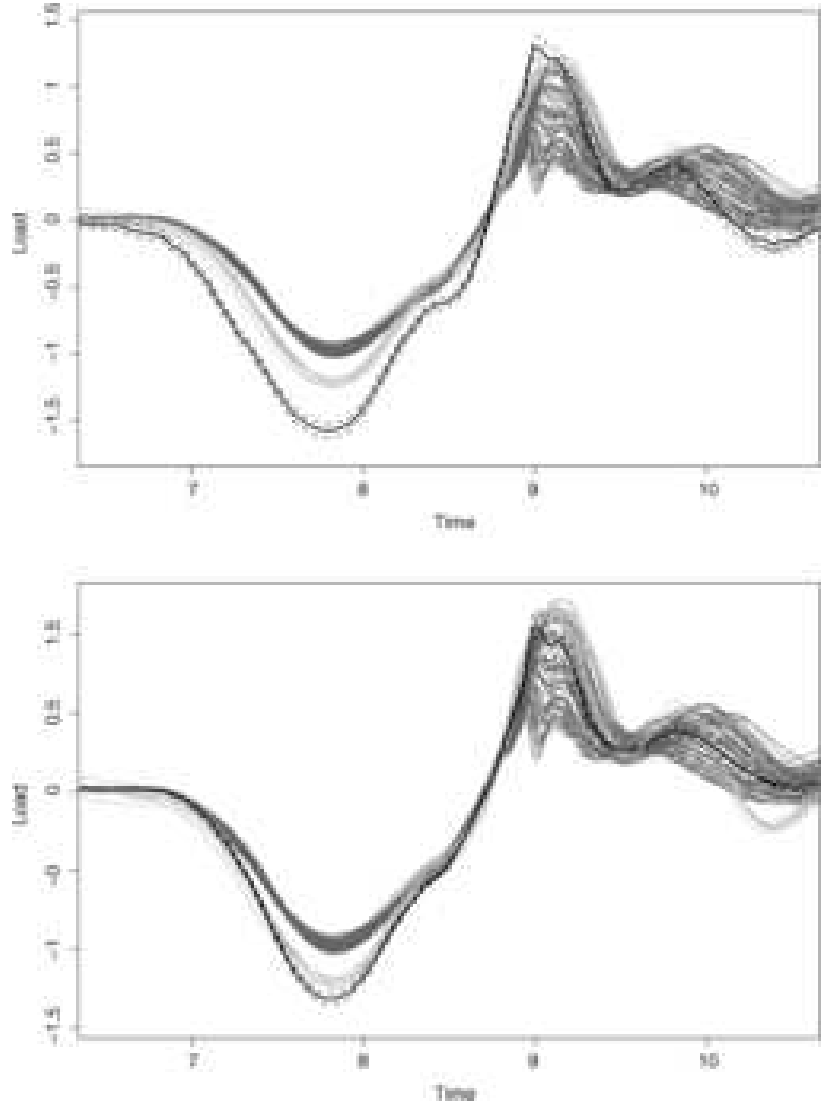

Fig. 10.   *Prediction at Site 2 of a new vehicle under Condition* B *in Region 1. Top: additive bias. Bottom: multiplicative bias. The solid black line is bias-corrected prediction, the dashed black lines are 90% tolerance bands and the grey lines are model runs; the light grey lines are the field runs later provided and not used in the analysis.*

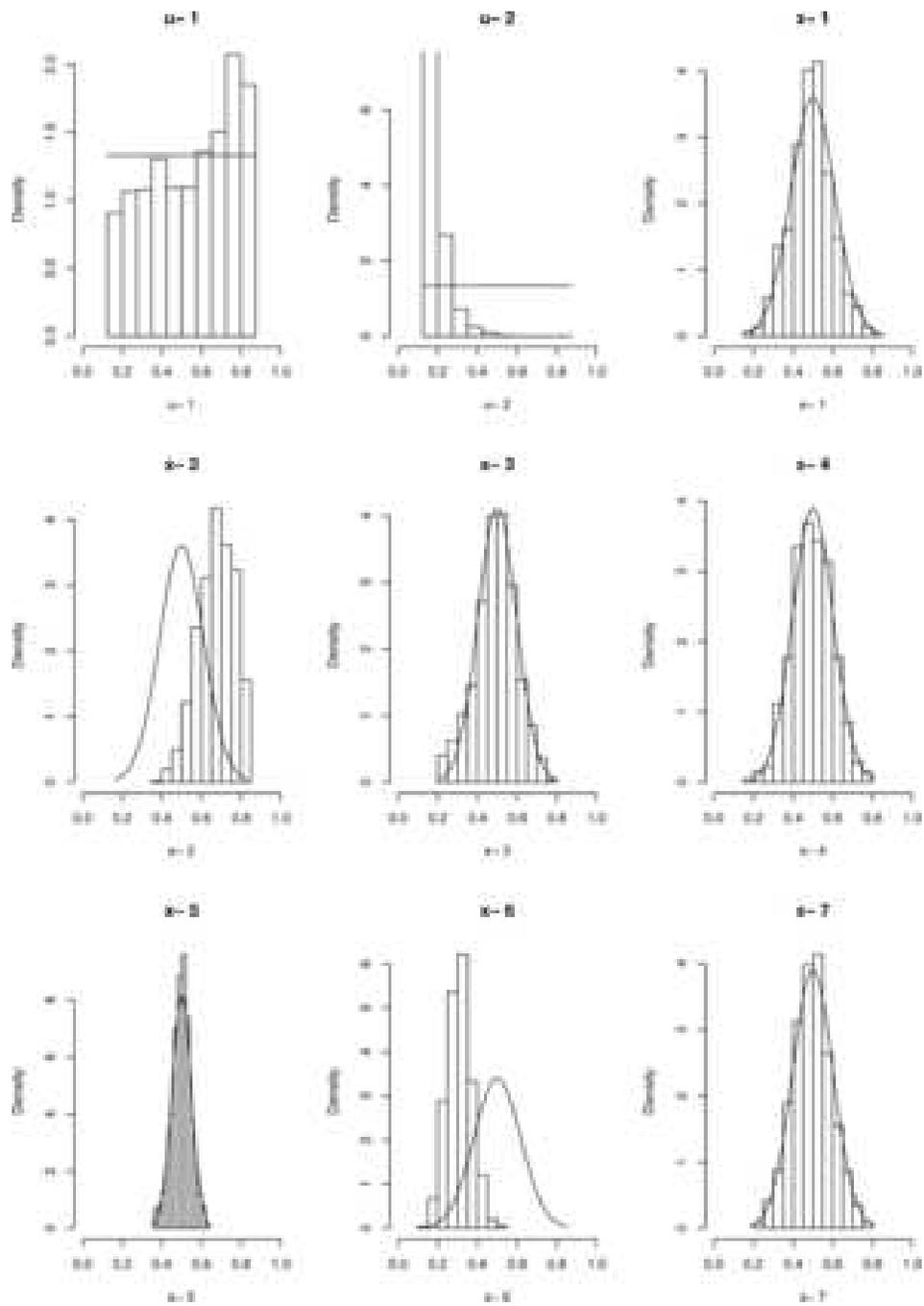

Fig. 11.   *Posterior distributions for the unknown input parameters for Site 2 of the system.*

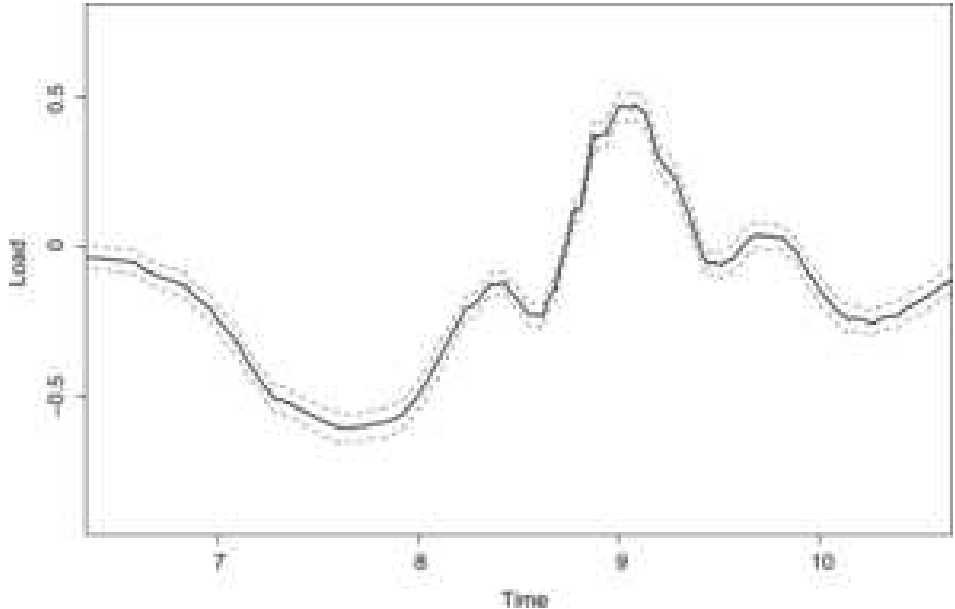

Fig. 12. *Bias estimate with 90% tolerance bounds for Site 2 at Region 1.*

Figure 12 presents the estimated bias function and Figure 13 the bias corrected and pure model predictions, along with the corresponding tolerance bounds. The presence of bias can be clearly seen, as well as the noticeably improved performance of the bias corrected prediction over the pure model prediction. Other figures for Site 2 are omitted since, with the exception of Figure 10, they do not provide further insight.

## APPENDIX A: DATA REGISTRATION AND WAVELET DECOMPOSITION

For the wavelet representations of the output curves it is important that the same wavelet basis elements simultaneously represent the important features of all curves. In the test bed problem the heights of the peaks and valleys of the curves from the field data are of primary importance, but their locations are not the same across the curves, due to random fluctuations in the tests. Thus we first *align* the curves so that the major peaks and the major valleys occur at the same locations. In other applications, alignment would likely be based on other key features of the curves. (In the test bed, the timing of the major events is not of concern—only the forces at these events are of interest. If it were important for the computer model to accurately reflect timing, as in the analysis of airbag deployment in Bayarri et al. [3], this mode of registration could not be used.)

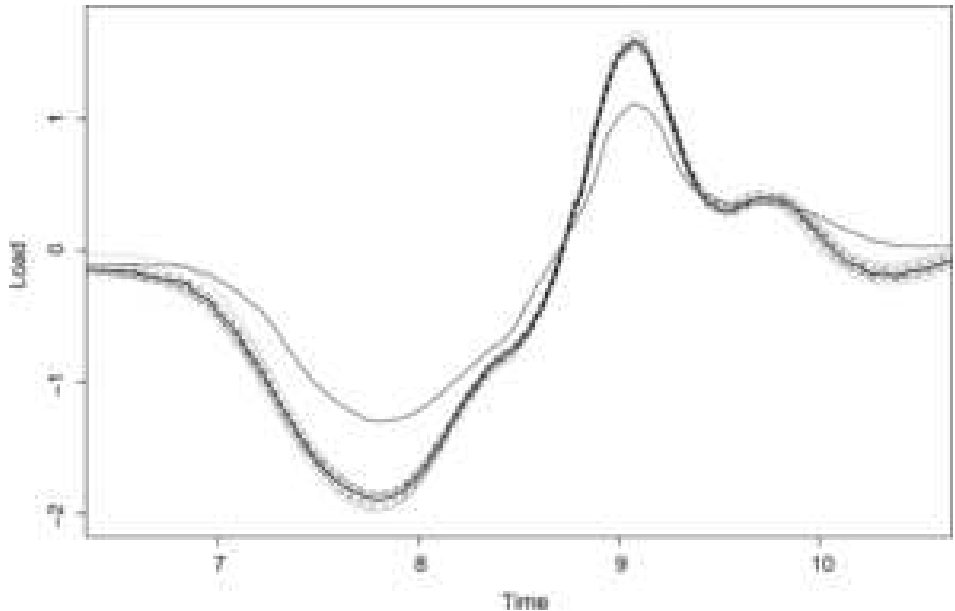

FIG. 13. *Bias-corrected prediction of reality with 90% tolerance bands for Site 2 at Region 1.*

We did not try to align the curves from the computer model runs, since variation in these curves could not be ascribed to random fluctuation. (One might think that the computer model curves would be automatically aligned but, surprisingly, in the test bed they did show some misalignment, perhaps due to differences in the damping parameters.) We construct a reference curve (for alignment of the field curves) by averaging the model-run curves and use piecewise linear transformation to align the peaks and valleys of the field curves to this reference curve. The details are as follows:

Step 1. Construct a dyadic grid (points of the form $j/2^q$ on the interval where the function is defined; for the test bed, the interval $[0, 65]$ covered the range of importance and $q = 12$). For each computer run extract the output values on the dyadic grid. Construct a pseudooutput for points not in the grid by linear interpolation and treat them as actual outputs.

Step 2. From the $K$ computer runs ($K = 65$ in the test bed) define the reference curve as $\bar{y}^M(t) = \frac{1}{K}\sum_{k=1}^{K} y^M(\mathbf{x}_k, \mathbf{u}_k; t)$.

- For the first major event, located in the region $6 < t < 11$, define
  - $A$ = location (value of $t$) of the maximum of the reference curve $\bar{y}^M(t)$;
  - $A_r^F$ = location of the maximum of $y_r^F$;

    – $a$ = location of minimum of $\bar{y}^M(t)$;
    – $a_r^F$ = location of minimum of $y_r^F$.
  • For the second major event, located in the region $37 < t < 41$,
    define $B, B_r^F, b, b_r^F$ analogously. Assume $a < A < b < B$.

Step 3.  For each $r$, match $a_r^F, A_r^F$ with $a, A$ by transforming $t$ in $[a_r^F, A_r^F]$ to

$$t' = a + (t - a_r^F)\frac{A - a}{A_r^F - a_r^F}.$$

Now define the registered $y_r^F$ on the interval $[a, A]$ as *registered $y_r^F(t')$ = original $y_r^F(t)$*, where $t'$ and $t$ are connected as above.

Step 4.  Assume that $b < B$. As in Step 3, register $y_r^F$ on the interval $[A_r^F, b_r^F]$ by mapping $[A_r^F, b_r^F]$ into $[A, b]$ via $t' = A + (t - A_r^F)\frac{b - A}{b_r^F - A_r^F}$. Similar registrations are done for the intervals $[0, a_r^F]$, $[b_r^F, B_r^F]$ and $[B_r^F, 65]$.

Figure 1 shows the registered data for Site 1 of the suspension system. "Time" in the figures is not literal time, but is a convenient scaling of the realigned time.

The wavelet decompositions are described in part in Section 3.2. The basis functions are such that at "level" 0 there is a scaling constant and for level $j = 1, \ldots, 12$ there are $2^{j-1}$ basis functions. To balance the need for approximation accuracy with the need to minimize the number of terms for computational feasibility, we considered each model-run and field curve and retained all coefficients at levels 0 through 3; for levels $j > 3$, we retained those coefficients that, in magnitude, were among the upper 2.5% of *all* coefficients at *all* levels for the given function, according to the R *wavethresh* thresholding procedure. We then took the union of all resulting basis functions for all the model-run and field curves. For the test bed there were 231 retained elements for the output from Site 1 on the system, and 213 for the output from Site 2. The combined (from both sites) number of retained elements was 289 and we used these for all analyses. The indices attached to these 289 retained basis elements are denoted by $I$.

## APPENDIX B: THE MCMC ALGORITHM

Step 1.  For $h = 1, \ldots, 1000$, sample the $\sigma_i^{2h}$ from the distribution

$$\text{InverseGamma}\left(3, \frac{2}{s_i^2}\right), \qquad \left(\text{shape} = 3, \text{scale} = \frac{2}{s_i^2}\right).$$

Step 2.  For $h = 1, \ldots, 1000$, make draws $\boldsymbol{\delta}^{*h}, \mathbf{u}^{*h}, \boldsymbol{\tau}^{2h}$ from the posterior distribution in (18). (This is complicated—the process is described last.)

Step 3.  Given $\boldsymbol{\delta}^{*h}, \mathbf{u}^{*h}, \boldsymbol{\sigma}^{2h}, \boldsymbol{\tau}^{2h}$, draw $\mathbf{w}^{bh}$ from the distribution in (17). (This is simply done by making a draw, for each $i$, from a normal distribution with the specified means and variances.)

Step 4. Given $\boldsymbol{\delta}^{*h}, \mathbf{u}^{*h}, \boldsymbol{\sigma}^{2h}, \boldsymbol{\tau}^{2h}, \mathbf{w}^{bh}$, make a draw of $w^{Mh}$ from the distribution in (16). (Again this is simply done by draws from normal distributions.)

For Step 2, we use a Metropolis–Hastings scheme to generate the $(h+1)$st sample. We break this up into two substeps.

Step 2.1. Propose $\boldsymbol{\tau}^2$ by generating from $q(\boldsymbol{\tau}^2 \mid \boldsymbol{\tau}^{2h}) = \prod_{i=0}^{12} q_i(\tau_i^2 \mid \tau_i^{2h})$, where

$$(28) \qquad q_i(\tau_i^2 \mid \tau_i^{2h}) \propto \begin{cases} \dfrac{1}{\tau_i^2}, & \text{if } \tau^2 \in [\tau_i^{2h} e^{-0.7}, \tau_i^{2h} e^{0.7}], \\ 0, & \text{otherwise.} \end{cases}$$

The posterior density of $\boldsymbol{\tau}^2$ is not very spiked, so this type of fairly broad local proposal works well. Finally, form the Metropolis–Hastings ratio

$$\rho = \frac{\pi(\boldsymbol{\delta}^h, \mathbf{u}^h, \boldsymbol{\tau}^2 \mid \boldsymbol{\sigma}^{2h}, \mathbf{D}) \, q(\boldsymbol{\tau}^{2h} \mid \boldsymbol{\tau}^2)}{\pi(\boldsymbol{\delta}^h, \mathbf{u}^h, \boldsymbol{\tau}^{2h} \mid \boldsymbol{\sigma}^{2h}, \mathbf{D}) \, q(\boldsymbol{\tau}^2 \mid \boldsymbol{\tau}^{2h})}$$

and define $\boldsymbol{\tau}^{2(h+1)} = \boldsymbol{\tau}^2$ with probability $\min(1, \rho)$; $\boldsymbol{\tau}^{2(h+1)} = \boldsymbol{\tau}^{2h}$ otherwise.

Step 2.2. Let $T_k^\delta = [a_k^\delta, A_k^\delta]$ and $T_k^u = [a_k^u, A_k^u]$ denote the intervals on which the prior densities for the corresponding variables are nonzero, and define

$$T_k^{*\delta h} = [\max(a_k^\delta, \delta_k^h - 0.05), \min(A_k^\delta, \delta_k^h + 0.05)],$$
$$T_k^{*uh} = [\max(a_k^u, u_k^h - 0.05), \min(A_k^u, u_k^h + 0.05)].$$

Propose $\boldsymbol{\delta}$, $\mathbf{u}$ from

$$g(\boldsymbol{\delta}, \mathbf{u} \mid \boldsymbol{\delta}^h, \mathbf{u}^h)$$
$$= \prod_{k=1}^{7} (\tfrac{1}{2} U(\delta_k \mid T_k^\delta) + \tfrac{1}{2} U(\delta_k \mid T_k^{*\delta h}))$$
$$\times \prod_{k=1}^{2} (\tfrac{1}{2} U(u_k \mid T_k^u) + \tfrac{1}{2} U(u_k \mid T_k^{*uh})).$$

The logic here is that the posterior densities for some of the parameters are quite flat, so that sampling uniformly over their support ($T_k^\delta$ or $T_k^u$) would be quite reasonable as a proposal. On the other hand, some of the posteriors are quite concentrated, and for these it is effective to use a locally uniform proposal, centered around the previous value and with a maximum step of 0.05; this leads to uniforms on $T_k^{*\delta h}$ or $T_k^{*uh}$, which are the regions defined by the intersection of the local uniforms and the support of the priors. Since the goal here was to create a procedure that can be automatically applied for this type of problem, 50–50 mixtures of the two proposals were adopted.

Finally, form the Metropolis–Hastings ratio

$$\rho = \frac{\pi(\boldsymbol{\delta}, \mathbf{u}, \boldsymbol{\tau}^{2(h+1)} \mid \boldsymbol{\sigma}^{2h}, \mathbf{D}) \, g(\boldsymbol{\delta}^h, \mathbf{u}^h \mid \boldsymbol{\delta}, \mathbf{u})}{\pi(\boldsymbol{\delta}^h, \mathbf{u}^h, \boldsymbol{\tau}^{2(h+1)} \mid \boldsymbol{\sigma}^{2h}, \mathbf{D}) \, g(\boldsymbol{\delta}, \mathbf{u} \mid \boldsymbol{\delta}^h, \mathbf{u}^h)}$$

and set $(\boldsymbol{\delta}^{(h+1)}, \mathbf{u}^{(h+1)}) = (\boldsymbol{\delta}, \mathbf{u})$ with probability $\min(1, \rho)$, and equal to $(\boldsymbol{\delta}^h, \mathbf{u}^h)$ otherwise.

These Metropolis–Hastings steps typically yield highly correlated iterations, so we actually cycle through them 200 times (with fixed $\boldsymbol{\sigma}^{2h}$) before saving the variable values for feeding into Steps 3 and 4.

**Acknowledgment.** Any opinions, findings and conclusions or recommendations expressed in this publication are entirely those of the authors and do not necessarily reflect the views of the National Science Foundation or the Spanish Ministry of Education.

M. J. Bayarri
Department of Statistics
  and Operations Research
Universitat de Valencia
Av. Dr. Moliner 50, 46100
Burjassot, Valencia
Spain
E-mail: susie.bayarri@uv.es

J. Cafeo
R. J. Parthasarathy
Research and Development
General Motors Corporation
30500 Mound Road
Warren, Michigan 48090
USA
E-mail: john.cafeo@gm.com

J. Palomo
Statistics and Operations
  Research Department
Rey Juan Carlos University
C/ Cuba, 33, 28945
Fuenlabrada (Madrid)
Spain
E-mail: jesus.palomo@urjc.es

J. Sacks
National Institute
  of Statistical Sciences
P.O. Box 14006
Research Triangle Park, North Carolina 27709
USA
E-mail: sacks@niss.org

J. O. Berger
F. Liu
Department of Statistical Science
Duke University
Durham, North Carolina 27708
USA
E-mail: berger@samsi.info
         fei@stat.duke.edu

G. Garcia-Donato
Department of Economy
Universidad de Castilla-La Mancha
Pza. Universidad, 2
02071 Albacete
Spain
E-mail: Gonzalo.GarciaDonato@uclm.es

R. Paulo
Departamento de Matematica
ISEG, Technical University of Lisbon
Rua do Quelhas, 6
1200-781 Lisboa
Portugal
E-mail: rui@iseg.utl.pt

D. Walsh
Institute of Information
  and Mathematical Sciences
Massey University
Auckland, Albany
New Zealand
E-mail: d.c.walsh@massey.ac.nz